\documentclass[twocolappendix,iop]{emulateapj}
\usepackage[utf8]{inputenc}
\usepackage{amsmath}
\usepackage{amssymb}
\usepackage{verbatim}
\usepackage{natbib}
\usepackage{graphicx,subfigure}   
\usepackage[colorlinks,urlcolor=blue,citecolor=blue,linkcolor=blue]{hyperref}
\renewcommand{\b}[1]{\boldsymbol{#1}} 

\newcommand{\bmath}[1]{\mbox{\boldmath{$#1$}}}

\newcommand{\bi}{\begin{itemize}}
\newcommand{\ei}{\end{itemize}}

\newcommand{\rli}{\,r_{\rm Li}}

\newcommand{\eq}[1]{Equation \eqref{eq:#1}}
\newcommand{\xsh}{x-x_{\rm sh}}
\newcommand{\xshnorm}{(\xsh)/\rli}
\newcommand{\eqn}[1]{(\ref{eq:#1})}
\newcommand{\fig}[1]{Figure~\ref{fig:#1}}
\newcommand{\fign}[1]{\ref{fig:#1}}

\newcommand{\sect}[1]{Section \ref{sec:#1}}
\newcommand{\app}[1]{Appendix \ref{sec:#1}}

\newcommand{\be}{\begin{eqnarray}}
\newcommand{\ee}{\end{eqnarray}}
\newcommand{\qt}{(1+q\,t)}

\newcommand{\perpe}{{e,\perp}}
\newcommand{\pare}{{e,\parallel}}
\newcommand{\perpi}{{i,\perp}}
\newcommand{\pari}{{i,\parallel}}
\newcommand{\alf}{Alfv\'en}

\def\L{\bmath{{L}}}
\def\bvec{\bmath{B}}

\newcommand{\Teperp}{T_{e,\perp} }

\begin{document}
\title{Electron Heating in Low Mach Number Perpendicular shocks. \\II. Dependence on the Pre-Shock Conditions}
\author{Xinyi Guo,$^1$ Lorenzo Sironi,$^2$ and Ramesh Narayan$^1$}
\affil{$^1$Harvard-Smithsonian Center for Astrophysics, 
60 Garden Street, Cambridge, MA 02138, USA
\\
$^2$Department of Astronomy, Columbia University, 550 W 120th St, New York, NY 10027, USA}
\email{xinyi.guo@cfa.harvard.edu}
 \email{lsironi@astro.columbia.edu}
 \email{rnarayan@cfa.harvard.edu}
\begin{abstract} Recent X-ray observations of merger shocks in galaxy clusters have shown that the post-shock plasma is two-temperature, with the protons being hotter than the electrons. In this work, the second of a series, we investigate  by means of two-dimensional particle-in-cell simulations the efficiency of electron irreversible heating in perpendicular low Mach number shocks. We consider values of plasma beta (ratio of thermal and magnetic pressures) in the range $4\lesssim \beta_{p0}\lesssim 32$ and sonic Mach number (ratio of shock speed to pre-shock sound speed) in the range $2\lesssim M_{s}\lesssim 5$, as appropriate for galaxy cluster shocks. As shown in Paper I, magnetic field amplification --- induced by shock compression of the pre-shock field, or by strong proton cyclotron and mirror modes accompanying the relaxation of proton temperature anisotropy --- can drive the electron temperature anisotropy beyond the threshold of the electron whistler instability. The growth of whistler waves breaks the electron adiabatic invariance, and allows for efficient entropy production. We find that the post-shock electron temperature $T_{e2}$ exceeds the adiabatic expectation $T_{e2,\rm ad}$ by an amount $(T_{e2}-T_{e2,\rm ad})/T_{e0}\simeq 0.044 \,M_s (M_s-1)$ (here, $T_{e0}$ is the pre-shock temperature), which depends only weakly on the plasma beta, over the range $4\lesssim \beta_{p0}\lesssim 32$ which we have explored, and on the proton-to-electron mass ratio (the coefficient of $\simeq 0.044$ is measured for our fiducial $m_i/m_e=49$, and we estimate that it will decrease to $\simeq 0.03$ for the realistic mass ratio). Our results have important implications for current and future observations of galaxy cluster shocks in the radio band (synchrotron emission and Sunyaev-Zel'dovich effect) and at X-ray frequencies.
\end{abstract}

\keywords{galaxies: clusters: general ---  instabilities --- radiation mechanisms: thermal --- shock waves}

\maketitle

\section{Introduction} \label{sec:intro}
Cluster merger shocks ---  collisionless low Mach number  shocks ($M_s \lesssim 5$, where $M_s$ is the ratio of shock speed to pre-shock sound speed) generated by infalling subclusters --- are routinely observed in the radio and X-ray bands.
X-ray measurements can quantify the density and temperature jumps between the unshocked (upstream) and the shocked (downstream) plasma \citep[e.g.,][]{Markevitch2002a,Finoguenov2010,Russell2010,Ogrean2013b,Eckert2016,Akamatsu2017}. The existence of shock-accelerated electrons is revealed by radio observations of synchrotron radiation \citep[e.g.,][]{VanWeeren2010,Lindner2014,Trasatti2015,Kale2017}.
Recently, the pressure jump associated with a merger shock has been measured through radio observations of the thermal Sunyaev-Zel'dovich (SZ) effect \citep{Basu2016}. 

Since all observational diagnostics are based on radiation emitted by electrons,  the proton properties (in particular, their temperature) are basically unconstrained. 
One usually makes the simplifying assumption that the electron temperature equals the proton temperature (and so, the mean gas temperature). This assumption is unlikely to hold in the vicinity of merger shocks, since most of the pre-shock energy is carried by protons and there is no obvious reason why protons should efficiently share with electrons the thermal energy they gain in passing through the shock.\footnote{Shocks in supernova remnants and the heliosphere are known to be two-temperature, with protons hotter than electrons \citep[e.g.,][]{Ghavamian2013}.}
While Coulomb collisions will eventually drive electrons and protons to equal temperatures,  the collisional equilibration timescale  \citep{Spitzer1962} for typical  conditions in the intracluster medium (ICM) is as long as  $10^{8}-10^9$ yrs. 
In fact, X-ray observations by \cite{Russell2012a} have shown that the electron temperature just behind a merger shock in Abell 2146 is lower than the mean gas temperature expected from the  Rankine-Hugoniot jump conditions, and thus lower than the proton temperature. On the other hand, \citet{markevitch_06} found that the temperatures across the shock in 1E~0657-56 (the so-called ``Bullet cluster'') are consistent with instant shock-heating of the electrons.

What is the mechanism responsible for electron heating at collisionless shocks, and how does the heating efficiency depend on the pre-shock conditions? This fundamental  question can be answered only through a detailed plasma physics analysis, since the fluid-type Rankine-Hugoniot relations only predict the jump in the mean plasma temperature across the shock, without specifying how the shock-generated heat is distributed between the two species. To understand electron heating in collisionless shocks, fully-kinetic simulations with the particle-in-cell (PIC) method \citep{Birdsall1991,Hockney1981} are essential to self-consistently capture  the role of electron and proton plasma instabilities in particle heating.

So far, most PIC studies of electron heating in shocks have focused on the regime of high sonic Mach number ($M_s\gtrsim10$, where $M_s$ is the ratio of the upstream flow speed relative to the shock to the upstream sound speed) and low plasma beta ($\beta_{p0}\lesssim 1$, where $\beta_{p0}$ is the ratio of thermal and magnetic pressures)  appropriate for supernova remnants \citep{Dieckmann2012,Matsukiyo2003,Matsukiyo2010}. In the first paper of this series \citep[][hereafter, Paper I]{guo17}, we investigated by means of analytical theory and two-dimensional (2D) PIC simulations  the physics of electron heating in low Mach number perpendicular shocks, a regime so far unexplored. As we summarize in \sect{phys}, we found that, while most of the electron heating is {\it adiabatic} --- induced by shock-compression of the upstream magnetic field --- the passage of electrons through the shock 
is also accompanied by entropy increase, i.e., by the production of {\it irreversible} electron heating. In analogy to the so-called ``magnetic pumping'' mechanism \citep{Spitzer1953,Berger1958,Borovsky1986}, we found that two basic ingredients are needed for electron irreversible heating: (\textit{i}) the presence of a temperature anisotropy, induced by field amplification coupled to adiabatic invariance; and (\textit{ii}) a mechanism to break the adiabatic invariance. We found that the growth of whistler waves --- triggered by the electron temperature anisotropy induced by field amplification --- was responsible for the violation of adiabatic invariance, and efficient entropy production. 


In Paper I, we validated our model for a  shock with Mach number $M_s=3$ and plasma beta $\beta_{p0}=16$, which we took as representative of merger shocks in galaxy clusters. In this work, we extend our investigation to a wide range of plasma beta ($4\lesssim \beta_{p0}\lesssim 32$) and sonic Mach number ($2\lesssim M_s\lesssim 5$). We quantify how the efficiency of electron heating and the post-shock electron-to-proton temperature ratio depend on $M_s$ and $\beta_{p0}$. We focus on perpendicular shocks (i.e., where the pre-shock field is orthogonal to the shock direction of propagation). The choice of a perpendicular magnetic field geometry is meant to minimize the role of non-thermal electrons, which are self-consistently accelerated in oblique configurations, as we have shown in  \citet{Guo2014,Guo2014c}. Because of the absence of shock-accelerated electrons returning upstream, the shock can settle down to a steady state on a shorter time, thus allowing us to focus on the steady-state electron heating physics. However,  we expect that the results presented in this paper will also apply to quasi-perpendicular configurations, as long as the non-thermal electrons are energetically sub-dominant. 

We find that the dependence on $M_s$ of the electron irreversible heating efficiency can be cast in a simple form: the post-shock electron temperature $T_{e2}$ exceeds the adiabatic expectation $T_{e2,\rm ad}$ by an amount that scales with Mach number as $(T_{e2}-T_{e2,\rm ad})/T_{e0}\simeq 0.044 \,M_s (M_s-1)$, where $T_{e0}$ is the pre-shock temperature. This depends only weakly on plasma beta (in the regime $4\lesssim \beta_{p0}\lesssim 32$ explored in this work) and on the proton-to-electron mass ratio (which we vary from 49 to 200).

The rest of the paper is organized as follows. In \sect{phys}, we summarize the results of Paper I, where we found that the field amplification required for efficient entropy production can be induced either by shock compression of the upstream field, or by growth of proton cyclotron and mirror modes accompanying the relaxation of proton temperature anisotropy.
With periodic box experiments meant to reproduce these two scenarios,  Section \ref{sec:ramp} (Section \ref{sec:waves}, respectively) investigates the dependence of the electron heating efficiency on Mach number and plasma beta, in a controlled setup where only the first mechanism (the second, respectively) is allowed to operate. The reader primarily interested in the implications of our study for shocks can skip Sections \ref{sec:ramp} and \ref{sec:waves} and proceed directly to Section \ref{sec:shock}, where we explore how the degree of electron irreversible heating depends on $M_s$ and $\beta_{p0}$, in full shock simulations (where the two processes discussed above generally co-exist). We present our key findings in \sect{disc} and conclude with a summary in Section \ref{sec:summ}.

\section{The Physics of Electron Heating}\label{sec:phys}
In this section, we summarize the main results of Paper I. As electrons pass through the shock, they experience a density compression, which results in adiabatic heating. In addition, irreversible processes operate, which  further increase the electron temperature. In Paper I, we found that efficient entropy production relies on the presence of two basic ingredients: (\textit{i}) a temperature anisotropy; and (\textit{ii}) a mechanism to break adiabatic invariance.
The change in electron entropy can then be written in two equivalent forms as:\footnote{Our analysis applies equally to electrons and protons, even though below we focus only on electron entropy.}
\begin{align}
ds_e & = &\left[\frac{1}{2}d\ln\left(\frac{T_{e,\parallel}}{(n/B)^2}\right)\right]\cdot\left[1-\frac{T_{e,\parallel}}{T_{e,\perp}}\right]-\frac{d e_{w, e}}{T_{e,\perp}}\,,\label{eq:dselec}\\
ds_e &= & -\left[d\ln \left(\frac{T_{e,\perp}}{B}\right)\right]\cdot\left[\frac{T_{e,\perp}}{T_{e,\parallel}}-1\right]-\frac{d e_{w, e}}{T_{e,\parallel}}\,,\label{eq:dsperp}
\end{align}
where $n$ is the electron density, $B$ the large-scale magnetic field strength (by ``large-scale,'' we mean the magnetic field on scales much larger than the electron Larmor radius and at frequencies much lower than the electron gyration frequency), and $T_\pare$ and $T_\perpe$ are the electron temperature parallel and perpendicular to the local magnetic field. The term $de_{w,e}$ on the right hand side of Equations \eqref{eq:dselec} and \eqref{eq:dsperp} represents the total energy per particle transferred to waves, including magnetic, electric and bulk kinetic contributions (in practice, we found that the magnetic term always dominates).\footnote{If the equations above were to be applied to protons, the corresponding term should include not only the energy residing in proton-driven waves, but also the energy lost by performing work on the electron plasma (see Paper I).} 

Note that the CGL double adiabatic theory of \citet{Chew1956} predicts that, for adiabatic perturbations,  $T_{e,\perp} \propto B$ and $T_{e,\parallel}\propto (n/B)^2$, which follow from the conservation of the first and second adiabatic invariants. The first square bracket on the right hand side of Equations \eqref{eq:dselec} and \eqref{eq:dsperp} explicitly shows that a mechanism to break the adiabatic invariance is needed for entropy production. In most cases, it is the temperature anisotropy (see the second square bracket on the right hand side of Equations \eqref{eq:dselec} and \eqref{eq:dsperp}) that provides the free energy for generating the waves responsible for breaking the adiabatic invariance.

\vspace{0.3in}
\subsection{Our Reference Shock}
In Paper I, we validated the heating model summarized above by performing PIC simulations with the electromagnetic PIC code TRISTAN-MP \citep{Buneman1993, Spitkovsky2005} for a representative shock with sonic Mach number $M_s=3$ and plasma beta $\beta_{p0}=16$. The Mach number 
\be\label{eq:machn}
M_s=\frac{V_1}{c_s}=\frac{V_1}{\sqrt{2 \Gamma k_{\rm B}T_0/m_i}}~
\ee
is defined as the ratio between the upstream flow velocity $V_1$ in the shock frame and the
upstream sound speed $c_s=\sqrt{2 \Gamma k_{\rm B}T_0/m_i}$. Here, $T_0$ is the upstream temperature (the same for both species, so $T_{e0}=T_{i0}=T_0$), $k_{\rm B}$ is the Boltzmann constant, $\Gamma=5/3$ is
the adiabatic index for an isotropic non-relativistic gas, and $m_i$
is the proton mass. The upstream 
 magnetic field strength  is parametrized by the plasma beta 
\begin{equation}\label{eq:plasmab}
\beta_{p0}=  \frac{8\pi n_{0} k_{\rm B} (T_{i0}+T_{e0})}{B_0^2}=\frac{16\pi n_{0} k_{\rm B} T_0}{B_0^2}~~,
\end{equation}
where $n_{i0}= n_{e0}= n_0$ is the number density of the incoming protons and electrons. Alternatively, one could quantify the magnetic field strength via the Alfv\'enic Mach number $M_A=M_s \sqrt{\Gamma \beta_{p0}/2}$. 

In Paper I, we considered a reference perpendicular shock with $M_s=3$ and $\beta_{p0}=16$ and showed that equations (\ref{eq:dselec}) and (\ref{eq:dsperp}) were in excellent agreement with the measured increase in electron entropy per particle (or specific entropy) across the shock, which can be computed directly from the electron distribution function $f_e({\b p})$ as
\begin{equation}\label{eq:entre}
s_e\equiv -\frac{\int d^3 p\, f_e\ln f_e}{\int d^3 p\, f_e}~. 
\end{equation}
In the reference shock, we found that efficient electron entropy production occurs at two major sites: at the shock ramp, where density compression coupled to flux freezing leads to field amplification (we call this scenario ``case A''); and farther downstream, where long-wavelength magnetic waves (more specifically, proton cyclotron and mirror modes) accompanying the relaxation of the temperature anisotropy of post-shock protons can also contribute to magnetic field growth (we call this scenario ``case B''). At both locations, field amplification coupled to adiabatic invariance drives the electrons to a large degree of temperature anisotropy, exceeding the threshold of the electron whistler instability. The resulting electron whistler waves --- whose presence is one of the common denominators at the two sites mentioned above --- cause efficient pitch angle scattering, which leads to violation of the electron adiabatic invariance and allows for  entropy increase.

In Paper I, we studied case A and case B in detail by employing controlled periodic box experiments meant to reproduce the shock conditions at the two major sites of entropy production. In particular, the shock physics in the ramp (case A) can be replicated in a periodic box where the PIC equations are modified to allow for a continuous large-scale compression, as in \citet{Sironi2015,Sironi2015a}. While in Paper I we studied this scenario only for our reference case with $M_s=3$ and $\beta_{p0}=16$, in \sect{ramp} of the present paper we extensively explore a range of Mach numbers and plasma betas.
In Paper I we also investigated the physics of electron heating via anisotropy-driven proton waves (case B) by means of  a periodic box initialized with anisotropic protons, with a degree of anisotropy inspired by our reference case with $M_s=3$ and $\beta_{p0}=16$. In \sect{waves}, we extend the same analysis to a wide range of flow conditions. 

The advantage of the periodic domains is twofold: (\textit{i}) they allow for more direct control of the relevant physics; and (\textit{ii}) due to less demanding computational requirements, they permit to extend our investigation up to the realistic mass ratio. In Paper I we were able to ascertain that the electron entropy increase has only a weak dependence on mass ratio (less than a $\sim 30\%$ drop, as we increase the mass ratio from $m_i/m_e=49$ up to $m_i/m_e=1600$).

\section{Electron Heating by Shock-Compression of the Upstream Field}\label{sec:ramp}
In this section, we focus on case A, i.e., we investigate the efficiency of electron heating (and its dependence on the flow conditions) when the field amplification that induces the electron anisotropy --- which in turn leads to electron whistler waves, and then to entropy increase --- is due exclusively to the large-scale density compression occurring in the shock ramp.

\subsection{Simulation Setup}
The simulation setup parallels the one employed in Section 5 of Paper I, which we summarize here for completeness.
We set up a suite of compressing box experiments, using the method introduced in \citet{Sironi2015,Sironi2015a}, that re-defined the unit length of the axes such that a particle subject only to compression stays at fixed coordinates in the primed system. Then, compression with rate $q$ is accounted for by the diagonal matrix 
\be\label{eq:eqL}
\L=\frac{\partial \bmath{x}}{\partial \bmath{x}'}=
\left(
\begin{array}{ccc}
\qt^{-1} \,\,\,& 0 \,& 0 \\
 0\,\,\,& 1 \,& 0 \\
 0 \,\,\,& 0 \,&1\\
\end{array}\right)~~~,
\ee
which has been tailored for compression along the $x$ axis,
perpendicular to the uniform ordered magnetic field $\bmath{B}_0$ initialized along the $y$ direction (in analogy to the shock setup that we will discuss in \sect{shock}).  Maxwell's equations  in the primed coordinate system automatically account for flux freezing (i.e., the field grows in time as $\bvec_0\qt$, in the same way as the density $n=n_0\qt$), and the form of the Lorentz force in the primed system guarantees the conservation of the first and second adiabatic invariants.

Since in Paper I we have shown that the wavevector of the whistler mode is nearly aligned with the field direction (i.e., along $\hat{y}$), we employ 1D simulations with the computational box oriented along $y$.  Yet, all three components of electromagnetic fields and particle velocities are tracked.  In 1D simulations, we can employ a large number of particles per cell (we use 1600 particles per species per cell) so we have adequate statistics for the calculation of the electron specific entropy from the phase space distribution function, as in \eq{entre}.

As a result of the large-scale compression encoded in \eq{eqL}, both electrons and protons will develop a temperature anisotropy, and we should witness the development of both electron and proton anisotropy-driven modes. However, as we have done in Paper I, our goal is to isolate the role of the large-scale field amplification (as expected in the shock ramp) in generating electron irreversible heating, regardless of the presence of proton-driven modes (which will be the focus of \sect{waves}). For this reason, in our compressing box runs, we artificially inhibit the update of the proton momentum (effectively, this corresponds to the case of infinitely massive protons). A similar strategy, but in the case of field amplification driven by shear, rather than compression, has been employed by \citet{riquelme_17}.

\begin{figure}
\begin{center}
\includegraphics[width=0.45\textwidth]{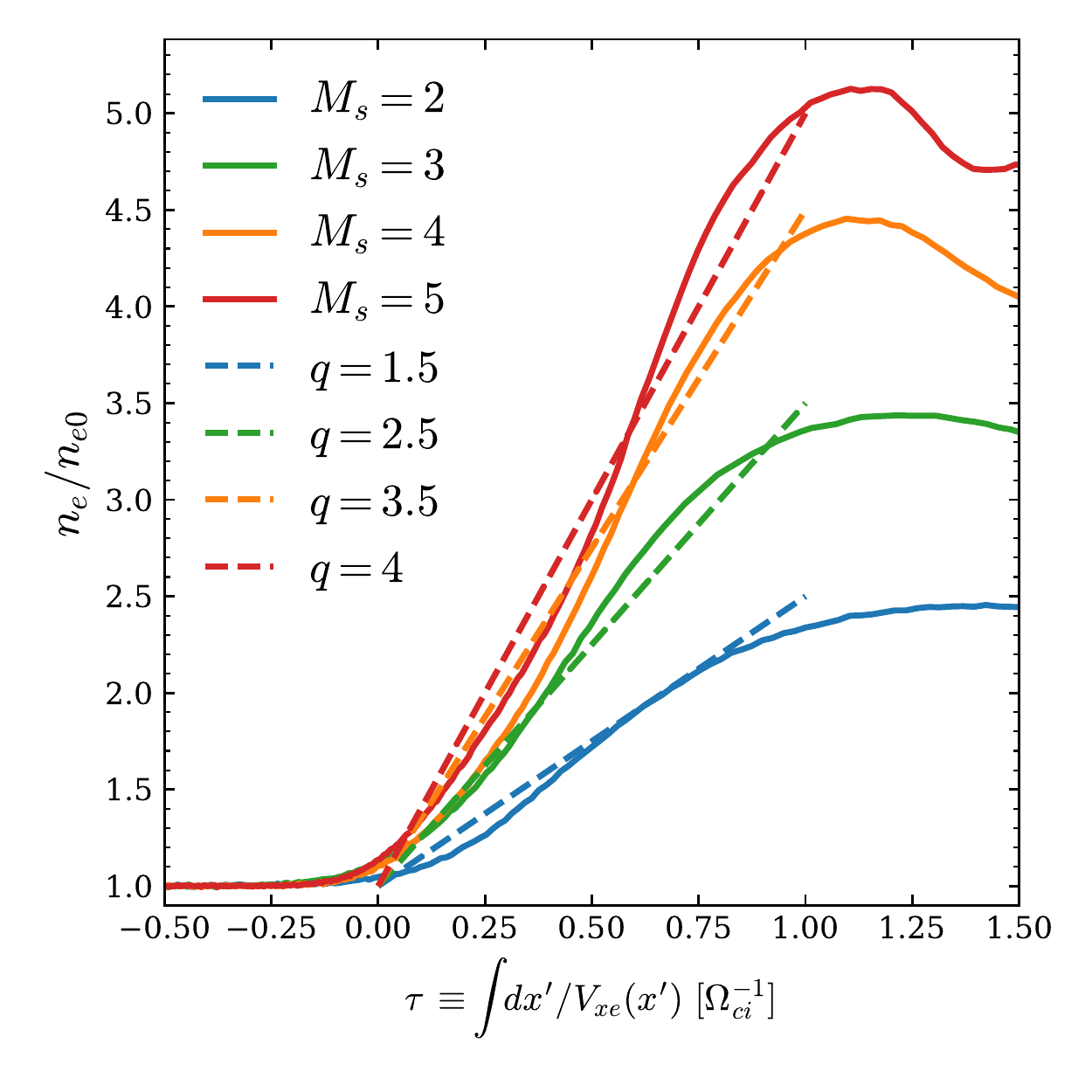}
\end{center}
\caption{Density compression experienced by electrons (solid lines) as they  propagate from upstream to downstream, as a function of the electron comoving time, from our suite of shock simulations (see \sect{shock}) with fixed $\beta_{p0}=16$ and varying $M_s$ (as indicated in the legend). Dotted lines show the linear approximation $n_e/n_{e0}=(1+qt)$ (with values of $q$ indicated in the legend) employed in the compressing box experiments.}\label{fig:compress_rate_ms}
\end{figure}

The compression rate $q$ (which we cast in units of $\Omega_{ci}=eB_0/m_i c$, i.e., the proton Larmor frequency in the initial field $B_0$) is measured directly from our  shock simulations described in \sect{shock}.
In fact, we can quantify the profile of electron density as a function of the co-moving time of the electron fluid
\begin{equation}\label{eq:electime}
\tau \equiv \int \frac {dx'}{V_{xe}(x')}
\end{equation}
where $V_{xe}$ is the electron fluid velocity in the shock  frame, and the integral goes from the upstream to the downstream region.\footnote{As described in \sect{shock}, the shock propagates along $+\hat{x}$ in our simulations.} 
Figure \ref{fig:compress_rate_ms} shows the electron density profile as a function of $\tau$ for a suite of shock simulations with $\beta_{p0}=16$ and varying Mach number (as indicated in the legend, solid lines), which will be discussed in \sect{shock}. The faster rise seen for higher values of $M_s$ is driven by the fact that the density jump across the shock monotonically increases with Mach number, for two reasons. First, the density jump from upstream to downstream as derived from the Rankine-Hugoniot relations is a monotonically increasing function of $M_s$. Second, the density overshoot in the shock ramp also increases with $M_s$ \citep[][see also \fig{shock_Mss} in \sect{shock}]{Leroy1983}. Since the thickness of the shock is nearly independent of Mach number (and always of order of the proton Larmor radius), a larger density jump for higher $M_s$ corresponds in \fig{compress_rate_ms} to a faster compression rate. In fact, if we model the density compression as a linear function of time (dashed lines in \fig{compress_rate_ms}), the resulting compression rate $q$  steadily increases with $M_s$. The values of $q$ adopted in the periodic simulations discussed in this section are given in Table \ref{table:compressbox} in units of the proton Larmor frequency $\Omega_{ci}$ (all the runs presented in this section employ a reduced mass ratio $m_i/m_e=49$). We remark that the value of the shock Mach number enters the compressing box experiments presented here only via the compression rate $q$ (i.e., $M_s$ should be meant as the Mach number of the  shock simulation that corresponds to a given choice of $q$ for the compressing box).

We summarize our physical and numerical parameters in Table \ref{table:compressbox} (the run names have the suffix ``c'' to indicate that we employ compressing boxes). The first four runs explore the dependence on Mach number (or equivalently, compression rate) for fixed $\beta_{p0}=16$, whereas the last four simulations investigate the dependence on $\beta_{p0}$ for a fixed compression rate $q=2.5\,\Omega_{ci}$, as appropriate for a shock with $M_s=3$. In all the runs, we initialize a population of isotropic electrons with temperature $T_{e0} = 10^{-2}m_e c^2/k_{\rm B}$. We resolve the electron skin depth
\be\label{eq:compe}
\frac{c}{\omega_{pe}} = \sqrt{\frac{m_e c^2}{4\pi e^2n_0}}~
\ee
with 10 cells, so the Debye length is marginally resolved. 
The box extent along the $y$ direction is fixed at $86\ c/\omega_{pe}$ for $\beta_{p0}\leq16$, which is sufficient to capture several wavelengths of the electron whistler  instability. For $\beta_{p0}=32$ and 64, we increase the box length roughly as $\propto\sqrt{\beta_{p0}}$, since the wavelength of  whistler waves increases with plasma beta (see \app{ewlinear}, where we study the linear dispersion properties of the whistler instability).

\begin{table}
\begin{center}
\begin{tabular}{ccccccc}
\hline 
run name & $M_{s}$ & $q$ & $\beta_{p0}$ & $k_{{\rm B}}T_{e0}/m_{e}c^{2}$ & $L_y\ [c/\omega_{pe}]$\tabularnewline
\hline 
\hline 
$\mathtt{Ms2c}$ & $2$ & $1.5$ & $16$ & $10^{-2}$ &  $86$\tabularnewline
$\mathtt{Ms3c/beta16c}$ & $3$ & $2.5$ & $16$ & $10^{-2}$ &   $86$\tabularnewline
$\mathtt{Ms4c}$ & $4$ & $3.5$ & $16$ & $10^{-2}$ &   $86$\tabularnewline
$\mathtt{Ms5c}$ & $5$ & $4$ & $16$ & $10^{-2}$ & $86$\tabularnewline
$\mathtt{beta4c}$ & $3$ & $2.5$ & $4$ & $10^{-2}$ &   $86$\tabularnewline
$\mathtt{beta8c}$ & $3$ & $2.5$ & $8$ & $10^{-2}$ &   $86$\tabularnewline
$\mathtt{beta32c}$ & $3$ & $2.5$ & $32$ & $10^{-2}$ &   $130$\tabularnewline
$\mathtt{beta64c}$ & $3$ & $2.5$ & $64$ & $10^{-2}$ &  $173$\tabularnewline
\hline 
\end{tabular}
\caption{Parameters for the compressing box experiments described in Section \ref{sec:ramp}. The compression rate $q$ is in units of the proton Larmor frequency $\Omega_{ci}$, for mass ratio $m_i/m_e=49$.}\label{table:compressbox}
\end{center}
\end{table}


\begin{figure}
\begin{center}
\includegraphics[width=0.4\textwidth]{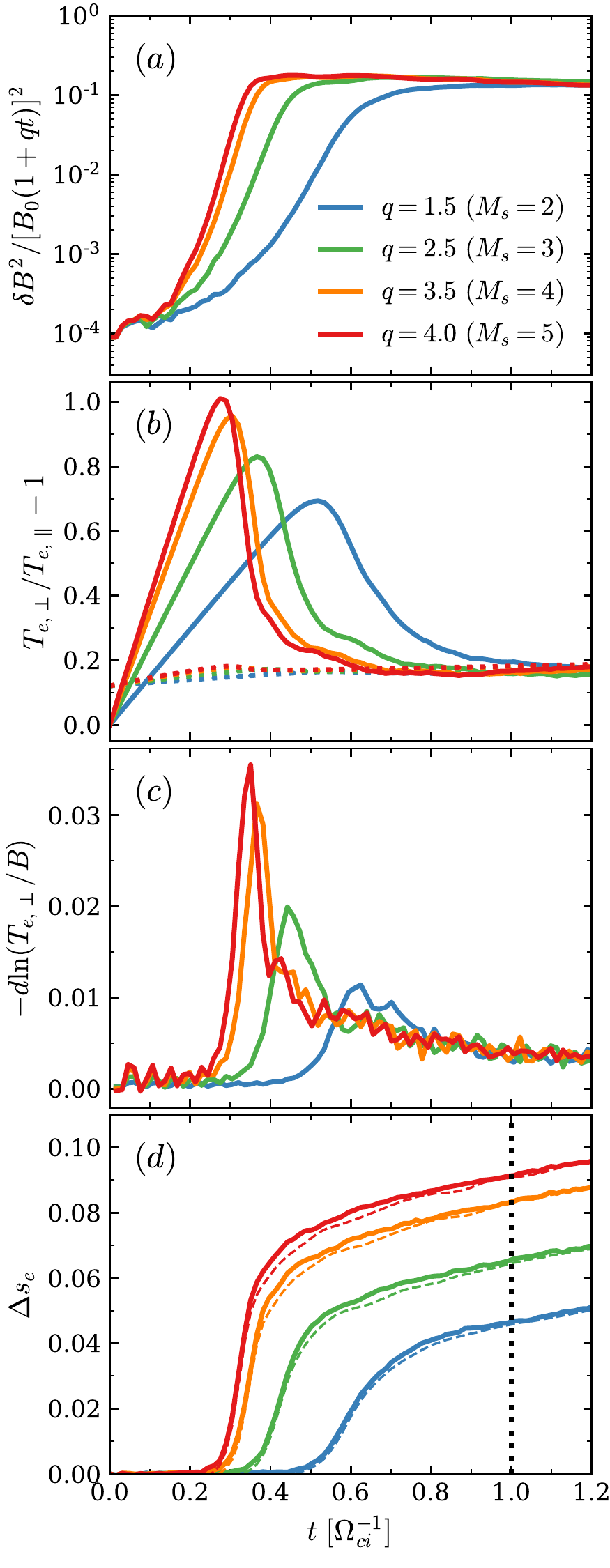}
\end{center}
\caption{Dependence on $M_{s}$, or equivalently compression rate $q$, of various space-averaged quantities in compressing box experiments $\mathtt{Ms2c}$, $\mathtt{Ms3c}$, $\mathtt{Ms4c}$ and $\mathtt{Ms5c}$, at fixed $\beta_{p0}$.  As a function of time in units of $\Omega_{ci}^{-1}$, we plot: (a) energy in magnetic field fluctuations, normalized to the energy of the compressed  field; (b) electron temperature anisotropy (solid lines) and threshold condition for the electron whistler instability (dotted lines with the same color coding as the solid lines); (c) rate of violation of adiabatic invariance $-d\ln (T_{e,\perp}/B)$; (d) electron entropy change, measured from the electron distribution function as in Equation \eqref{eq:entre} (solid lines) and predicted from our heating model of \eq{dselec} (thin dashed lines). The vertical dotted black line in panel (d) marks the approximate end of the compression phase in the shock ramp. 
}\label{fig:compressbox_Ms}
\end{figure}

\subsection{Dependence on $M_{s}$}\label{sec:rampM}
Figure \ref{fig:compressbox_Ms} compares the results of compressing box simulations (runs $\mathtt{Ms2c}$, $\mathtt{Ms3c}$, $\mathtt{Ms4c}$, $\mathtt{Ms5c}$ in Table \ref{table:compressbox}) with the same $\beta_{p0}=16$ and different compression rates $q$ (or equivalently, different Mach numbers of the corresponding shock simulations). We present the evolution of the whistler wave energy (panel (a)), the electron temperature anisotropy (panel (b)), the rate $-d\ln (T_\perpe/B)$ of breaking adiabatic invariance (panel (c)) and the electron entropy increase (panel (d)) when varying the compression rate from $q/\Omega_{ci}=1.5$ up to 4.0 (from blue to red, see the legend in the first panel). For mass ratio $m_i/m_e=49$, the compression rate in units of the electron gyration frequency $\Omega_{ce}=(m_i/m_e)\Omega_{ci}$ is  $q/\Omega_{ce}\simeq 0.02\,(q/\Omega_{ci})\ll1$, i.e., compression occurs slowly as compared to the electron gyration time (so, our choice of $m_i/m_e=49$ fulfills the requirement $q/\Omega_{ce}\ll1$ expected for the realistic mass ratio). 

As a result of the large-scale compression, the electron perpendicular and parallel temperatures are expected to scale as $\Teperp\propto B\propto (1+qt)$ and $T_{e,\parallel}\propto (n/B)^2\propto \;$const, according to the double adiabatic theory. In fact, the electron anisotropy at early times grows as $T_\perpe/T_\pare-1= qt$ (Figure \ref{fig:compressbox_Ms}(b)), thus, at a faster rate for higher $q$ (or equivalently, in higher $M_{s}$ shocks).

The increasing temperature anisotropy leads to the exponential growth of the electron whistler instability. When the energy in whistler waves reaches a fraction $\sim 10^{-2}$ of the compressed background field energy (\fig{compressbox_Ms}(a)), the waves are 
sufficiently strong to scatter the electrons in pitch angle, breaking their adiabatic invariance and decreasing the electron anisotropy. In fact, the peak in panel (c), i.e., the time when the electron adiabatic invariance is most violently broken, always corresponds to the time when the electron anisotropy in panel (b) shows the sharpest decrease. This occurs earlier for higher $q$ (if time is measured in $\Omega_{ci}^{-1}$, as in \fig{compressbox_Ms}), since electrons are driven sooner to large levels of anisotropy.
As a result of efficient pitch angle scattering, the electron anisotropy is reduced to the marginal stability threshold of the electron
 whistler instability \citep{Gary2005}
\begin{equation}\label{eq:threshwhis}
\frac{T_{e,\perp}}{T_{e,\parallel}}-1 \simeq \frac{ 0.21 }{\beta_{e,\parallel}^{0.6}}
\end{equation}
(dotted lines in Figure \ref{fig:compressbox_Ms}(b), with the same color coding as in the legend of panel (a)), which is nearly the same for all the runs, as they start with the same $\beta_{p0}$ and maintain a similar value of $\beta_{e,\parallel}$. Here, $\beta_\pare$ is the electron plasma beta measured with the parallel temperature $T_\pare$.

Near the end of the exponential growth of whistler waves, the electron entropy shows a rapid increase (panel (d)). Here,   the electron anisotropy is still large, and at the same time whistler waves are sufficiently powerful to provide effective pitch-angle scattering. In other words, both terms in the square brackets of either \eq{dselec} or \eq{dsperp} are large. The resulting entropy increase is a monotonic function of $q$, for the following reason. First, larger values of $q$ allow the electrons to reach higher levels of peak anisotropy (\fig{compressbox_Ms}(b)). This can be understood from the competition between the large-scale compression rate (which increases the electron anisotropy) and the growth rate of whistler waves (that try to reduce the anisotropy via pitch angle scattering). Since the whistler growth rate depends on how much the anisotropy exceeds the whistler threshold in \eq{threshwhis}, a higher anisotropy is needed for larger $q$ (in fact, \fig{compressbox_Ms} shows that the whistler growth rate is higher for larger $q$). Second, since whistler waves are sourced by the free energy in electron anistropy, the wave energy at saturation will be larger for higher $q$ (Figure \ref{fig:compressbox_Ms}(a)). Third, stronger whistler waves will be more efficient in breaking the electron adiabatic invariance (Figure \ref{fig:compressbox_Ms}(c)). The combination of these effects explains the monotonic trend in electron entropy observed in \fig{compressbox_Ms}(d) near the end of the exponential growth of whistler waves (i.e., at the time of sharp increase in the curves of panel (d)). 

After the exponential growth, electron whistler waves enter a secular phase where the wave energy (normalized  to the compressed background  field energy) stays almost constant in time. At this point, the whistler wave energy is also nearly independent of $q$ (panel (a)), and the same will hold for the rate of violation of electron adiabatic invariance (panel (c)). The electron anisotropy settles around the threshold of marginal stability (compare solid and dotted lines in panel (b) at late times), which is independent of $q$ at fixed $\beta_\pare$. It follows that the rate of increase of electron entropy  during the secular phase will be the same regardless of $q$, as indeed confirmed by panel (d) (see the slow growth at late times, at a rate independent of $q$). We remark that the overall increase of electron entropy as measured directly from our simulations using \eq{entre} is in excellent agreement with our heating model of \eq{dselec} (compare solid and thin dashed lines in panel (d), respectively).


From \fig{compressbox_Ms}(d), we can infer how the entropy increase in the shock ramp should scale with Mach number, if field amplification is induced by shock-compression of the upstream field (case A). Since the compression in the shock ramp lasts about one proton gyration time, we compare the entropy curves at $\Omega_{ci}t\sim1$, as indicated by the vertical dotted black line in panel (d). We then find that the efficiency of electron heating in the shock ramp is expected to be larger at higher $M_s$ (corresponding to faster compressions in \fig{compressbox_Ms}), at fixed $\beta_{p0}$. We will confirm this trend in our shock simulations presented in \sect{shock}. However, we anticipate that in shocks with high Mach number, proton-driven waves already appear near the shock ramp. In this case, large-scale field compression and proton-driven waves (as discussed in \sect{waves}) co-exist and co-contribute to efficient electron entropy production in the shock transition region.

\begin{figure}
\begin{center}
\includegraphics[width=0.4\textwidth]{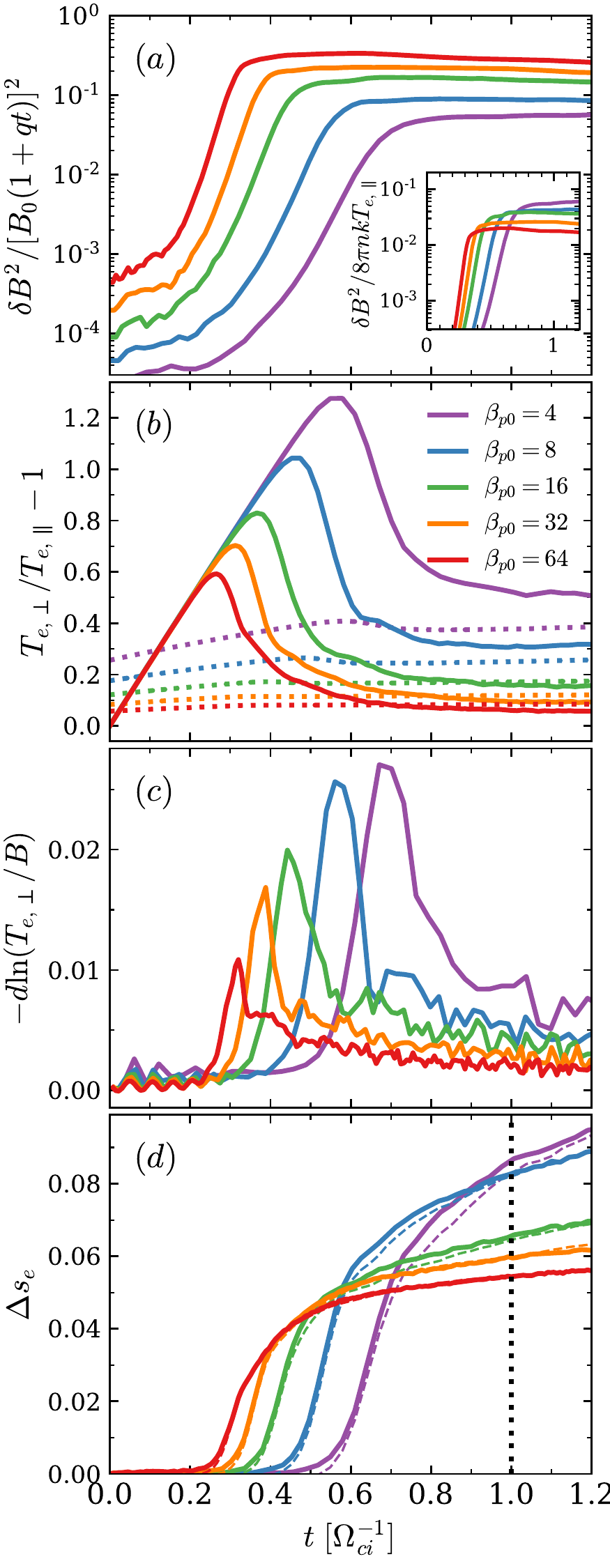}
\end{center}
\caption{Dependence on $\beta_{p0}$ of various space-averaged quantities in compressing box experiments $\mathtt{beta4c}$, $\mathtt{beta8c}$, $\mathtt{beta16c}$, $\mathtt{beta32c}$ and $\mathtt{beta64c}$, at fixed $M_s=3$.  
As a function of time in units of $\Omega_{ci}^{-1}$, we plot: (a) energy in magnetic field fluctuations, normalized to the energy of the compressed  field; in the inset we also plot the pressure in magnetic field fluctuations normalized by the electron parallel pressure; (b) electron temperature anisotropy (solid lines) and threshold condition for the electron whistler instability (dotted lines with the same color coding as the solid lines); (c) rate of violation of adiabatic invariance $-d\ln (T_{e,\perp}/B)$; (d) electron entropy change, measured from the electron distribution function as in Equation \eqref{eq:entre} (solid lines) and predicted from our heating model of \eq{dselec} (thin dashed lines). The vertical dotted black line in panel (d) marks the approximate end of the compression phase in the shock ramp. }\label{fig:compressbox_betap}
\end{figure}

\subsection{Dependence on $\beta_{p0}$}\label{sec:rampbeta}
In this subsection, we study the dependence of the electron heating efficiency on plasma beta $\beta_{p0}$, by means of compressing box experiments (as appropriate for case A). We fix the compression rate at $q=2.5\,\Omega_{ci}$, as appropriate for a shock with $M_s=3$, and vary the initial plasma beta $\beta_{p0}$ from $4$ to $64$ (runs $\mathtt{beta4c}$, $\mathtt{beta8c}$, $\mathtt{beta16c}$, $\mathtt{beta32c}$, $\mathtt{beta64c}$). 

Figure \ref{fig:compressbox_betap} compares the results of our runs. Initially, the electron temperature anisotropy increases  as $T_{e,\perp}/T_{\parallel}-1 = qt$, regardless of $\beta_{p0}$ (panel (b)). Eventually, this leads to the development of the whistler instability, whose exponential growth needs to balance the large-scale compression (and so, its growth rate is nearly independent of $\beta_{p0}$, as confirmed by panel (a)). As we discuss in \app{ewlinear}, a given growth rate of the whistler mode requires a lower degree of electron anisotropy for higher plasma beta, which explains the trend in peak anisotropy of \fig{compressbox_betap}(b). In turn, a lower level of peak anisotropy corresponds to a smaller amount of free energy available to be converted into whistler waves, whose amplitude is indeed weaker at higher $\beta_{p0}$ (see the inset in \fig{compressbox_betap}(a), where we normalize the whistler wave pressure with respect to the electron parallel pressure $n k_{\rm B} T_\pare$). Weaker whistler wave activity at higher $\beta_{p0}$ explains why the breaking of electron adiabatic invariance is less violent at higher beta (panel (c)). Since both the electron anisotropy and the violation of adiabatic invariance are weaker for larger $\beta_{p0}$, the resulting entropy increase near the end of the exponential phase of whistler growth is smaller for higher plasma beta, as shown in \fig{compressbox_betap}(d) (solid lines; compare the values attained during the early phase of rapid growth).

In the secular phase of the electron whistler instability, when efficient pitch angle scattering has brought the anisotropy down to the threshold of marginal stability (dotted lines in \fig{compressbox_betap}(b), with the same color coding as the solid lines), the electron entropy keeps increasing, yet at a slower rate. In the secular phase, the trend discussed above still holds (i.e., weaker entropy production at higher plasma beta), primarily because the degree of electron anisotropy stays smaller at higher $\beta_{p0}$ (panel (b)), due to the monotonic dependence on $\beta_\pare$ of the threshold of marginal stability, see \eq{threshwhis}.  As we have emphasized in the previous subsection, the increase of electron entropy in both exponential and secular phases, as measured directly from our simulations  (\eq{entre}), is in excellent agreement with our heating model of \eq{dselec} (compare solid and thin dashed lines in panel (d), respectively). 

In summary, when field amplification is due to compression alone (case A), the efficiency of electron entropy production decreases monotonically with increasing plasma beta. In particular, this holds after one proton gyration time (see the vertical dotted black line in panel (d)), which we have taken to be the characteristic compression time in the shock ramp. However, as well shall see in the next section, the dependence on $\beta_{p0}$ is opposite when field amplification is provided by proton-driven waves. As a result of the two opposite trends, in our shock simulations of \sect{shock}, where both compression and proton-driven waves contribute to field amplification, the dependence on $\beta_{p0}$ is rather weak, in the range of plasma beta that we explore.

\begin{table*}[!ht]
\centering
\begin{tabular}{ccccccccc}
\hline 
run name & $M_{s}$ & $\beta_{p0}$ & $k_{{\rm B}}T_{i0\rm box,\parallel}/m_{i}c^{2}$ & $T_{i0\rm box,\perp}/T_{i0\rm box,\parallel}$ & $T_{e0\rm box}/T_{i0\rm box,\parallel}$ & $\beta_{e0\rm box}$ & $N_{{\rm ppc}}$ & $L_y\ [c/\omega_{pi}]$\tabularnewline
\hline 
\hline 
$\mathtt{Ms2nc}$ & $2$ & $16$ & $2\times10^{-4}$ & $2.88$ & $1.64$ & $6.69$ & $10^{4}$ & $30.9$\tabularnewline
$\mathtt{Ms3nc/beta16nc}$ & $3$ & $16$ & $2\times10^{-4}$ & $6.87$ & $1.95$ & $6.43$ & $10^{4}$ & $30.9$\tabularnewline
$\mathtt{Ms4nc}$ & $4$ & $16$ & $2\times10^{-4}$ & $12.86$ & $2.10$ & $6.34$ & $10^{4}$ & $30.9$\tabularnewline
$\mathtt{Ms5nc}$ & $5$ & $16$ & $2\times10^{-4}$ & $20.74$ & $2.18$ & $6.30$ & $10^{4}$ & $30.9$\tabularnewline
$\mathtt{beta8nc}$ & $3$ & $8$ & $2\times10^{-4}$ & $6.69$ & $1.93$ & $3.22$ & $10^{4}$ & $30.9$\tabularnewline
$\mathtt{beta32nc}$ & $3$ & $32$ & $2\times10^{-4}$ & $6.96$ & $1.96$ & $12.85$ & $10^{4}$ & $30.9$\tabularnewline
$\mathtt{beta64nc}$ & $3$ & $64$ & $2\times10^{-4}$ & $7.00$ & $1.96$ & $25.69$ & $2\times10^{4}$ & $46.3$\tabularnewline
\hline 
\end{tabular}
\caption{Parameters for the anisotropic protons box experiments described in Section \ref{sec:waves}. We employ a fixed mass ratio $m_i/m_e=49$ and resolve the electron skin depth with 7 cells, which is sufficient to capture the electron Debye length.}\label{table:undrivenbox}
\end{table*}

\section{Electron Heating by Proton-Driven Waves}\label{sec:waves}
In this section, we focus on case B, i.e., we investigate the efficiency of  electron irreversible heating (and its dependence on the flow conditions) when  field amplification  is due exclusively to proton-scale waves induced by the relaxation of proton temperature anisotropy.
To study this effect, we employ periodic boxes (not compressing, so with $q=0$), where the degree of proton anisotropy is prescribed to mimic the conditions expected in the shock downstream. Since both the proton cyclotron instability, which dominates over the mirror mode in our  runs (see \sect{shock}), and the electron whistler instability have the fastest growing wavevector aligned with the background field, we employ 1D simulation domains aligned with the $y$ direction of the field.

\subsection{Simulation Setup}\label{sec:setupmodel}
We now describe how the initial conditions of our 1D periodic box experiments are set up to be representative of the downstream region of a shock  with Mach number $M_s$ and plasma beta $\beta_{p0}$. We need to prescribe the initial proton temperatures perpendicular and parallel to the magnetic field ($T_{i0\rm box,\perp}$ and $T_{i0\rm box,\parallel}$), the initial electron temperature ($T_{e0\rm box}$; as we justify below, we consider isotropic electrons) and the initial electron plasma beta ($\beta_{e0\rm box}$). We employ the subscript ``box'' to distinguish them from the initial conditions of our shock simulations. In principle, these parameters can be set by measuring directly the flow conditions behind our simulated shocks. However, we show below that they can be simply prescribed   using the Rankine-Hugoniot jump conditions. This has the advantage of showing explicitly the expected dependence on the shock Mach number $M_s$ and plasma beta $\beta_{p0}$.

Since the pre-shock flow moves along the $x$ direction, while the pre-shock magnetic field is oriented along $y$ (see \sect{shock}), the post-shock protons will be promptly gyrotropic in the plane $xz$ perpendicular to the pre-shock field, but generally anisotropic with respect to the $y$ direction parallel to the field. In the absence of instabilities that mediate efficient isotropization, the motions perpendicular and parallel to the field are effectively decoupled, and the post-shock protons behave as a plasma with two degrees of freedom (so, with adiabatic index $\Gamma=2$). This is demonstrated by the 2D out-of-plane and 1D shock simulations shown in Appendix A of Paper I. The density and temperature jumps across a shock with adiabatic index $\Gamma=2$ (as appropriate for the region just behind the shock) are 
\begin{equation}\label{eq:n2rh2}
r_{\rm RH, \Gamma=2} =\frac{3 \beta_{p0} M_s^2}{2 + \beta_{p0}\left( 2 + M_s^2\right)}~~,
\end{equation}
\begin{eqnarray}\label{eq:t2rh2}
\!\!\!\!\!\! \Delta t_{\rm RH, \Gamma=2} =
 M_s^2\!\left(\!2 - \frac{2}{r_{\rm RH,\Gamma=2}^2}\!\right)\! +\! \frac{4\!-\!4\,r_{\rm RH,\Gamma=2}}{\beta_{p0}}\! +\! 4~~,
\end{eqnarray}
where $\Delta t_{\rm RH, \Gamma=2} $ is, more precisely, the jump in the perpendicular temperature for the overall fluid (the parallel one stays unchanged).

We then assume  that the electron temperature jump in the perpendicular direction follows from the adiabatic law (in \sect{shock}, we quantify the efficiency of electron irreversible heating; still, for the parameter regime we explore, most of the electron temperature increase comes from adiabatic compression). Coupling electron adiabatic invariance with flux freezing, this implies that the post-shock electron temperature is $r_{\rm RH, \Gamma=2}\,T_0$. From \eq{t2rh2}, which prescribes the jump in perpendicular temperature for the overall (electron $+$ proton) fluid, we find that the post-shock proton temperature in the direction perpendicular to the field, which we take as the initial condition $T_{i0\rm box,\perp}$ in our periodic boxes, will be
\begin{equation}\label{eq:perpe2}
T_{i0\rm box,\perp} = T_{0} (2\,\Delta t_{\rm RH, \Gamma=2} -  r_{\rm RH, \Gamma=2})~. 
\end{equation}
The parallel proton and electron temperatures are the same as in the pre-shock region, and in particular
\begin{equation}\label{eq:pare2}
T_{i0\rm box,\parallel} = T_{0} ~.  
\end{equation}

As regard to the initialization of electrons, we assume that the electron anisotropy is rapidly erased due to the rapid development of electron-scale instabilities, as it indeed occurs in the shock ramp.\footnote{Despite the fact that we now assume isotropic electrons, the ansatz of a $\Gamma=2$ gas that we employed to calculate the jump conditions will still hold, since the post-shock pressure is mostly contributed by protons (see \sect{shock}).} If this happens without any energy exchange with the protons, the isotropic electron temperature is
\be
T_{e0\rm box}=T_0 \frac{2\,r_{\rm RH, \Gamma=2}+1}{3}~.
\ee
Finally, the initial electron plasma beta can be computed from flux freezing as
\begin{equation}
\beta_{e0\rm box} = \frac{2 \,r_{\rm RH, \Gamma=2}+1}{3\, r_{\rm RH, \Gamma=2}} \frac{\beta_{p0}}{2}~.
\end{equation}

The physical and numerical parameters used for the periodic box experiments of this section are summarized in 
Table \ref{table:undrivenbox} (the run names have the suffix ``nc'' to indicate that the boxes are not compressing). We indicate the Mach number $M_s$ and plasma beta $\beta_{p0}$ of the corresponding shock simulations, the initialization values of
$k_{{\rm B}}T_{i0\rm box,\parallel}/m_{i}c^{2}$,  $T_{i0\rm box,\perp}/T_{i0\rm box,\parallel}$, $T_{e0\rm box}/T_{i0\rm box,\parallel}$ and $\beta_{e0\rm box}$ employed in the periodic boxes, the number of computational particles per cell ($N_{\rm ppc}$) and the box length $L_y$ along the $y$ direction of the background large-scale field, in units of the proton skin depth $c/\omega_{pi}=\sqrt{m_i/m_e}\,c/\omega_{ pe}$. We employ a reduced mass ratio $m_i/m_e=49$. From Table \ref{table:undrivenbox}, it is apparent that the Mach number $M_s$ has a pronounced effect on the initial proton anisotropy. In fact, in the limit of $M_s\gg1$, we have $r_{\rm RH, \Gamma=2}\propto M_s^0$ and $\Delta t_{\rm RH, \Gamma=2}  \propto M_s^2$, so that  $T_{i0\rm box,\perp}/T_{i0\rm box,\parallel}\propto M_s^2$ from Equations \eqn{perpe2} and \eqn{pare2}.

\begin{figure}
\begin{center}
\includegraphics[width=0.37\textwidth]{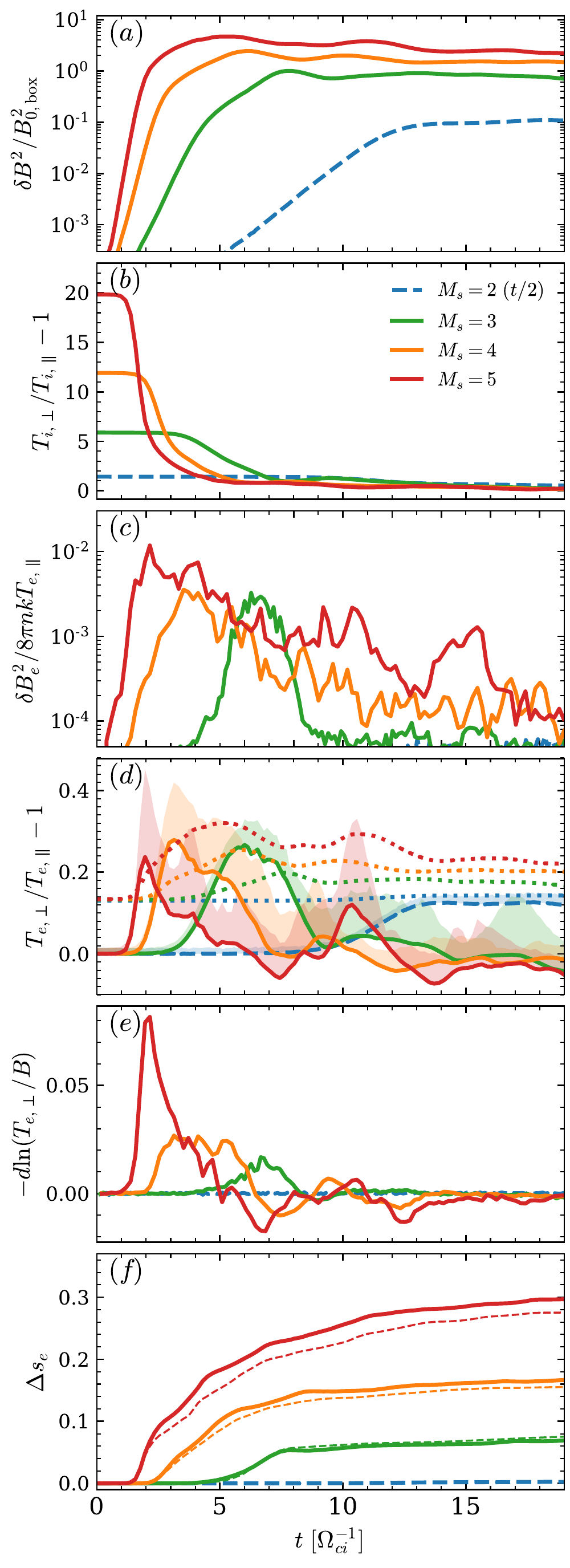}
\end{center}
\vspace{-.175in}
\caption{Dependence on $M_{s}$ of various quantities in periodic box experiments with anisotropic protons (runs $\mathtt{Ms2nc}$, $\mathtt{Ms3nc}$, $\mathtt{Ms4nc}$, $\mathtt{Ms5nc}$), at fixed $\beta_{p0}=16$. As a function of time in units of $\Omega_{ci}^{-1}$,  we plot: (a) energy in magnetic field fluctuations, normalized to the energy of the initial field in the box $B_{0,\rm box}$; (b) proton temperature anisotropy; (c) magnetic pressure in electron-scale fluctuations, normalized to the electron parallel pressure; (d) electron temperature anisotropy (solid lines for the box-averaged values, shaded regions for the $50\%-90\%$ percentile) and threshold of the electron whistler instability (dotted lines with the same color coding as the solid lines); (e) rate of violation of the electron adiabatic invariance $-d\ln (T_{e,\perp}/B)$; (f) electron entropy change, measured from the electron distribution function as in Equation \eqref{eq:entre} (solid lines) and predicted from our heating model of \eq{dselec} (thin dashed lines). 
The time axis of the 
$M_{s}=2$ run (blue dashed lines) is reduced by a factor of two for better comparison.
To isolate whistler-like waves in panel (c), we have applied a high-pass filter that retains frequencies higher than $0.064\,\Omega_{ce}$ and wavelengths shorter than $25\,c/\omega_{pe}$.}
\label{fig:undriven_Ms}
\end{figure}

\subsection{Dependence on $M_{s}$}\label{sec:waveM}
In this subsection, we compare the periodic box simulations $\mathtt{Ms2nc}$, $\mathtt{Ms3nc}$, $\mathtt{Ms4nc}$, $\mathtt{Ms5nc}$,
that correspond to shocks with fixed $\beta_{p0}=16$ but different $M_{s}$ ranging from $2$ to $5$. Our results are shown in Figure \ref{fig:undriven_Ms}. The case corresponding to $M_s=2$ is plotted with a dashed line to emphasize that its time axis has been reduced by a factor of two, to facilitate comparison with the other cases.

As we have discussed above (see also Table \ref{table:undrivenbox}), runs with larger $M_s$ are initialized with a stronger degree of proton anisotropy (see the curves in panel (b) at the initial time). In response to the greater amount of free energy stored in proton temperature anisotropy, runs with larger $M_s$ develop stronger proton cyclotron waves (panel (a)). In addition, since the growth rate of the proton cyclotron instability is higher for larger temperature anisotropies, at fixed plasma beta (see the linear dispersion properties of the proton cyclotron instability in \app{ICI}), the proton cyclotron wave energy grows faster at higher $M_s$ (panel (a)). Since pitch angle scattering by the proton cyclotron modes is responsible for relaxing the temperature anisotropy, this explains why the proton anisotropy in \fig{undriven_Ms}(b) drops faster for higher $M_s$, despite starting from higher initial values.

The growth of proton cyclotron waves provides a source of field amplification that can perform work on the electrons. However, this does not automatically lead to  electron irreversible heating. In fact, \fig{undriven_Ms} shows that for $M_s=2$ the energy in proton cyclotron modes is so small (dashed blue  line in panel (a)) that the electron temperature anisotropy (dashed blue line in panel (d)) never exceeds the threshold of the electron whistler instability (indicated in \fig{undriven_Ms}(d) by the corresponding dotted blue line). In fact, in this case no whistler waves are observed to grow (no blue line appears in \fig{undriven_Ms}(c), where we plot the magnetic pressure associated with whistler waves in units of the electron parallel pressure). In 
the absence of whistler waves, the electron evolution stays adiabatic (dashed blue line in panel (e)), and no electron entropy is generated (dashed blue line in panel (f)).
 
For higher $M_s$,  the proton waves are sufficiently strong to drive the electron anisotropy beyond the whistler threshold (compare solid and dotted lines of the same color in \fig{undriven_Ms}(d)). Since whistler waves provide the pitch-angle scattering required to break adiabatic invariance, this leads to electron entropy production (\fig{undriven_Ms}(f)). As shown in \fig{undriven_Ms}(f), the increase in electron entropy is a monotonic function of $M_s$. As we now discuss, this is related to the fact that the same trend holds separately for the two terms in square brackets of \eq{dselec} (or equivalently, \eq{dsperp}).

In fact, the degree of violation of electron adiabatic invariance is larger for higher $M_s$ (\fig{undriven_Ms}(e)), since whistler waves are more powerful (\fig{undriven_Ms}(c)).\footnote{In order to obtain the magnetic energy $\delta B_e^2/8 \pi$ in whistler-like fluctuations, we have applied a high-pass filter in frequency and wavenumber, as described in Paper I.} The trend in electron anisotropy (\fig{undriven_Ms}(d)) is less clear, with the green, orange and red curves showing comparable peak values. Unlike in the compressing box experiments of the previous section, where the degree of electron anisotropy was roughly uniform throughout the simulation domain, here proton-driven modes introduce large spatial variations in the electron properties (in analogy to Figure 10 of Paper I). It is then useful to complement the information on the box-averaged anisotropy (solid lines in \fig{undriven_Ms}(d)) with shaded regions indicating the $50\% - 90\%$ percentile of electron temperature anisotropy. This reveals that spatial variations in electron anisotropy are stronger at higher $M_s$ (or equivalently, the distance between the solid curve and the upper boundary of the corresponding shaded region increases with $M_s$), where proton modes are also more powerful. In addition, runs with larger $M_s$ present localized regions with systematically higher peak anisotropies (the peak of the shaded red region is higher than the orange one, which in turn is higher than the green one). On the one hand, this justifies the fact that the strength of whistler waves, which are seeded by electron anisotropy, increases with $M_s$. On the other hand, when coupled with the dependence on $M_s$ of the violation of adiabatic invariance (\fig{undriven_Ms}(e)), it fully justifies why runs with higher $M_s$ lead to more efficient electron entropy production. 

While for low $M_s$ most of the entropy increase occurs near the end of the exponential phase of the proton cyclotron instability, at high $M_s$ a steady growth in electron entropy is observed during the secular stage. Here, the strong cyclotron modes can occasionally excite local patches of electron anisotropy that exceed the whistler threshold (e.g., see the peak in the red shaded region at $\Omega_{ci}t\sim 10.5$ in \fig{undriven_Ms}(d)). The resulting whistler activity (see the corresponding peaks in the curves of panels (c) and (e)) can further increase the electron entropy.

Finally, we point out that, as we have also discussed in the case of compressing box simulations, the increase of electron entropy, as measured directly from our simulations  (\eq{entre}), is in excellent agreement with our heating model of \eq{dselec} (compare solid and thin dashed lines in panel (f), respectively).

\begin{figure}
\begin{center}
\includegraphics[width=0.41\textwidth]{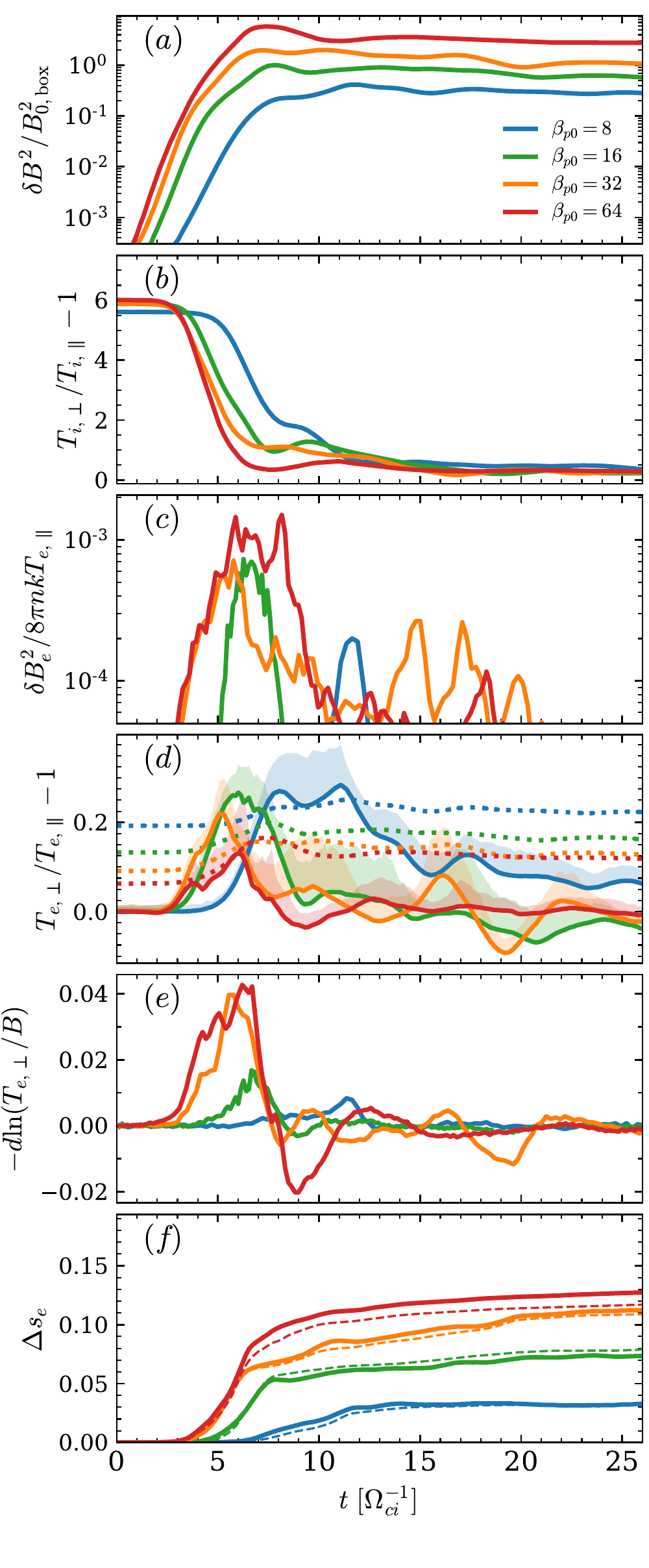}
\end{center}
\vspace{-.35in}
\caption{Dependence on $\beta_{p0}$ of various quantities in periodic box experiments  with anisotropic protons (runs $\mathtt{beta8nc}$, $\mathtt{beta16nc}$, $\mathtt{beta32nc}$, $\mathtt{beta64nc}$), at fixed $M_s=3$. 
As a function of time in units of $\Omega_{ci}^{-1}$, we plot: (a) energy in magnetic field fluctuations, normalized to the energy of the initial  field in the box $B_{0,\rm box}$; (b) proton temperature anisotropy; (c) magnetic pressure in electron-scale fluctuations, normalized to the electron parallel pressure; (d) electron temperature anisotropy (solid lines for the box-averaged values, shaded regions for the 50\% -- 90\% percentile) and threshold of the electron whistler instability (dotted lines with the same color coding as the solid lines); (e) rate of violation of the electron adiabatic invariance $-d\ln (T_{e,\perp}/B)$; (f) electron entropy change, measured from the electron distribution function as in Equation \eqref{eq:entre} (solid lines) and predicted from our heating model of \eq{dselec} (thin dashed lines). 
To isolate whistler-like waves in panel (c), we have applied a high-pass filter: for run  $\mathtt{beta8nc}$ (respectively, $\mathtt{beta16nc}$, $\mathtt{beta32nc}$, $\mathtt{beta64nc}$) we retain frequencies higher than $0.063\,\Omega_{ce}$ (respectively, $0.063$, $0.039$, $0.039$), and wavelengths shorter than $20\,c/\omega_{pe}$ (respectively, $25$, $35$, $35$).These choices are motivated by the fact that the fastest growing mode of the electron whistler instability has lower frequencies and longer wavelengths for higher plasma beta. }\label{fig:undriven_betap}
\end{figure}

\subsection{Dependence on $\beta_{p0}$}\label{sec:wavebeta}
In this subsection, we compare the periodic box simulations $\mathtt{beta8nc}$, $\mathtt{beta16nc}$, $\mathtt{beta32nc}$, $\mathtt{beta64nc}$,
that correspond to shocks with fixed $M_s=3$ but different $\beta_{p0}$ ranging from $8$ to $64$. Our results are shown in Figure \ref{fig:undriven_betap}. 

At fixed $M_s$, the initial proton temperature anisotropy is nearly independent of $\beta_{p0}$ (see \fig{undriven_betap}(b) at early times). From the dispersion properties of the proton cyclotron mode (see \app{ICI}), at fixed proton anisotropy the growth is faster at higher plasma beta, in agreement with the exponential phase in \fig{undriven_betap}(a) and with the drop in proton anisotropy in \fig{undriven_betap}(b). At fixed proton anisotropy, the free energy available to be converted into proton waves will be larger with increasing $\beta_{p0}$, just because the proton thermal content is higher. In fact, 
when normalized to the magnetic energy of the large-scale field $B_{0,\rm box}$, the magnetic energy of proton cyclotron waves at saturation is larger for higher plasma beta, as shown in \fig{undriven_betap}(a). 

In all the cases that we investigate, the growth of proton cyclotron waves drives the electron anisotropy above the whistler threshold (Figure \ref{fig:undriven_betap}(d), compare solid and dotted lines). At higher  $\beta_{p0}$, the electron whistler instability starts earlier (Figure \ref{fig:undriven_betap}(c)), due to the combination of two effects: the proton waves grow faster, and the whistler threshold is lower, so easier to be exceeded.

In addition, the stronger proton waves generated for higher $\beta_{p0}$ will have the chance to perform more work onto the electrons. In particular, at higher $\beta_{p0}$ we expect that electrons will be driven further into the unstable region of the electron whistler mode (i.e., beyond the marginal stability threshold, which is itself a function of plasma beta, see \eq{threshwhis}).
This is suggested by the general trend seen in \fig{undriven_betap}(d), where, e.g., the orange solid curve (for $\beta_{p0}=32$) lies at $\Omega_{ci}t\sim 5$ significantly above the corresponding marginal stability threshold (indicated by the dotted orange line), whereas the solid blue line (for $\beta_{p0}=8$) at $\Omega_{ci}t\sim 11$ is only marginally above the corresponding threshold (dotted blue curve). In turn,  a higher level of electron anisotropy (with respect to the baseline provided by the marginal stability threshold) results in stronger whistler wave activity (\fig{undriven_betap}(c)), and in more violent breaking of electron adiabatic invariance (\fig{undriven_betap}(e)). It follows that the electron entropy increase will be more pronounced for higher $\beta_{p0}$, as confirmed by \fig{undriven_betap}(f).

We conclude with two comments. First, once again, our heating model is in good agreement with the measured entropy increase (compare thin dashed and solid lines in \fig{undriven_betap}(f)). Second, while in the compressing box experiments of the previous section (meant to mimic field amplification by shock compression of the upstream field), higher values of plasma beta resulted in a  weaker increase in electron entropy, the opposite trend is observed here, where field amplification is due to proton cyclotron waves. In the shock simulations presented in the next section, where the two processes will co-exist, we should expect a weak dependence on plasma beta, as indeed we will find for the parameter regime we explore.

\section{Electron Heating in Shocks}\label{sec:shock}
In the previous two sections, we have investigated the efficiency of  electron irreversible heating when the magnetic field amplification that induces electron anisotropy --- which in turn sources the growth of whistler waves, and eventually results in entropy production --- is due to two separate mechanisms: in \sect{ramp}, we have discussed case A, where field amplification is due to a large-scale density compression; in \sect{waves}, we have focused on case B, where proton waves accompanying the relaxation of proton anisotropy can increase the magnetic field strength. In shocks, the two mechanisms co-exist, as we have already discussed in Paper I for our reference case with $M_s=3$ and $\beta_{p0}=16$. We now explore the dependence of irreversible electron heating in perpendicular shocks on Mach number and plasma beta. In \sect{disc}, we summarize the key findings from the shock simulations and provide an empirical fit to our results, which can be used in comparing with the observations.

\begin{table}
\centering
\begin{tabular}{ccccccc}
\hline 
run name & $M_{s}$ & $M_{s,{\rm meas}}$ & $\beta_{p0}$ & $m_{i}/m_{e}$ & $N_{{\rm ppc}}$ & $L_{y}\ [c/\omega_{pi}]$\tabularnewline
\hline 
\hline 
$\mathtt{Ms2beta4}$ & $2$ & $2.13$ & $4$ & $49$ & $32$ & $21.6$\tabularnewline
$\mathtt{Ms2beta8}$ & $2$ & $2.11$ & $8$ & $49$ & $32$ & $21.6$\tabularnewline
$\mathtt{Ms2beta16}$ & $2$ & $2.16$ & $16$ & $49$ & $32$ & $21.6$\tabularnewline
$\mathtt{Ms2beta32}$ & $2$ & $2.11$ & $32$ & $49$ & $32$ & $21.6$\tabularnewline
$\mathtt{Ms3beta4}$ & $3$ & $3.04$ & $4$ & $49$ & $32$ & $21.6$\tabularnewline
$\mathtt{Ms3beta8}$ & $3$ & $3.03$ & $8$ & $49$ & $32$ & $21.6$\tabularnewline
$\mathtt{Ms3beta16}$ & $3$ & $2.98$ & $16$ & $49$ & $32$ & $21.6$\tabularnewline
$\mathtt{Ms3beta32}$ & $3$ & $2.95$ & $32$ & $49$ & $32$ & $21.6$\tabularnewline
$\mathtt{Ms4beta4}$ & $4$ & $4.06$ & $4$ & $49$ & $32$ & $21.6$\tabularnewline
$\mathtt{Ms4beta8}$ & $4$ & $3.92$ & $8$ & $49$ & $32$ & $21.6$\tabularnewline
$\mathtt{Ms4beta16}$ & $4$ & $3.94$ & $16$ & $49$ & $32$ & $21.6$\tabularnewline
$\mathtt{Ms4beta32}$ & $4$ & $3.94$ & $32$ & $49$ & $32$ & $21.6$\tabularnewline
$\mathtt{Ms5beta4}$ & $5$ & $4.91$ & $4$ & $49$ & $32$ & $21.6$\tabularnewline
$\mathtt{Ms5beta8}$ & $5$ & $4.92$ & $8$ & $49$ & $32$ & $21.6$\tabularnewline
$\mathtt{Ms5beta16}$ & $5$ & $4.92$ & $16$ & $49$ & $32$ & $21.6$\tabularnewline
$\mathtt{Ms5beta32}$ & $5$ & $4.94$ & $32$ & $49$ & $32$ & $21.6$\tabularnewline
$\mathtt{mi200Ms2}$ & $2$ & $2.16$ & $16$ & $200$ & $48$ & $21.4$\tabularnewline
$\mathtt{mi200Ms3}$ & $3$ & $2.98$ & $16$ & $200$ & $64$ & $21.4$\tabularnewline
$\mathtt{mi200Ms4}$ & $4$ & $3.94$ & $16$ & $200$ & $48$ & $21.4$\tabularnewline
$\mathtt{mi200Ms5}$ & $5$ & $4.92$ & $16$ & $200$ & $48$ & $21.4$\tabularnewline
\hline 
\end{tabular}
\caption{Parameters for the shock simulations presented in \sect{shock}. All the runs are initialized with $T_{i0}=T_{e0}=T_{0}=10^{-2}m_{e}c^{2}/k_{{\rm B}}$ and $c/\omega_{pe}=10$ cells. }\label{table:shockruns}
\end{table}

\subsection{Simulation Setup}
We perform shock simulations using the 3D electromagnetic PIC code TRISTAN-MP \citep{Buneman1993,Spitkovsky2005}. Our setup parallels closely what we have employed in Paper I.
 We use a 2D simulation box in the $x-y$ plane, with periodic boundary conditions in the $y$ direction. Yet, all three components of particle velocities and electromagnetic fields are tracked. The shock is set up by reflecting an upstream electron-proton plasma moving along the $-\hat{x}$ direction off a conducting wall at the leftmost boundary of the computational box ($x=0$). The interplay between the reflected stream and the incoming plasma causes a shock to form, which propagates along $+\hat{x}$. In the simulation
frame, the downstream plasma is at rest. The pre-shock magnetic field is initialized along the $\hat{y}$ direction perpendicular to the shock direction of propagation (i.e., we focus on the case of a ``perpendicular'' shock). Our 2D setup allows to capture the physics of both electron (whistler) and proton (mirror and proton cyclotron) anisotropy-driven instabilities that are crucial for electron heating.

We vary  the shock Mach number $M_s$ (defined in \eq{machn}) and the pre-shock plasma beta $\beta_{p0}$ (defined in \eq{plasmab}) as listed in Table \ref{table:shockruns}, where we summarize the physical and numerical parameters of our shock simulations. The values of $M_s$ and $\beta_{p0}$ cover the regime of pre-shock conditions expected in galaxy cluster shocks.

We point out that, in setting up our simulations, we do not have direct control on the upstream velocity in the shock frame $V_1$ (which enters the definition of $M_s$), but only on the upstream velocity in the downstream  frame $V_0$. In other words, we have no direct way of prescribing $M_s$.  
In order to obtain a given ``target'' Mach number $M_s$, we iteratively solve the Rankine-Hugoniot jump conditions and select the value of $V_0$ that corresponds to the chosen $M_s$. Since in the parameter regime covered by our simulations, the protons in the immediate post-shock region retain a significant degree of anisotropy, we solve the Rankine-Hugoniot relations assuming a 2D adiabatic index $\Gamma=2$. It follows that the actual value of Mach number measured {\it a posteriori} in our simulations ($M_{s,\rm meas}$ in Table \ref{table:shockruns}) may differ from the Mach number $M_s$ targeted {\it a priori}. In practice,  Table \ref{table:shockruns} shows that the two values differ at most by a few percent.

The pre-shock particles are injected at a ``moving injector", which recedes from the wall in the $+\hat{x}$ direction at the speed of light. For further numerical optimization, we allow the moving injector to periodically jump backward (i.e. in the $-\hat{x}$ direction), so that its distance ahead of the shock is always of order of a few proton Larmor radii (see Paper I for details).
 The pre-shock particles are initialized as a drifting
Maxwell-J\"uttner distribution with a temperature $T_{i0}=T_{e0}=T_{0}=10^{-2}m_{e}c^{2}/k_{{\rm B}}$. We resolve the electron skin depth $c/\omega_{pe}$ (defined in \eq{compe}) with $10$ computational cells, so that the electron Debye length is appropriately captured. We use a time resolution of $dt = 0.045\ \omega_{pe}^{-1}$. The number of particles per cell $N_{\rm ppc}$ (including both species) is  in Table \ref{table:shockruns}. Convergence checks as regard to spatial resolution and number of particles per cell have been performed in Paper I.

The shock structure is controlled by the proton Larmor radius 
\begin{equation}\label{eq:rLi}
r_{{\rm Li}}=\frac{V_0}{v_A}\sqrt{\frac{m_i}{m_e}}\ \frac{c}{\omega_{pe}}\gg \frac{c}{\omega_{pe}}~,
\end{equation} 
where the \alf\ speed is $v_A=B_0/\sqrt{4 \pi m_i n_0}$. Similarly, the evolution of the 
shock occurs on a time scale given by the proton Larmor gyration period $\Omega_{ci}^{-1}=r_{\rm Li}V_0^{-1}\gg \omega_{pe}^{-1}$.
The need to resolve the electron scales, and at the same time to capture the shock evolution for many $\Omega_{ci}^{-1}$, is an enormous computational challenge for the realistic mass ratio $m_i/m_e=1836$. Thus we adopt a reduced mass ratio $m_i/m_e=49$ for most of our runs, but we have tested that a higher mass ratio yields identical results (in \app{mime}, we explore the dependence on Mach number of a few simulations with $m_i/m_e=200$). In Paper I, we have argued that for our reference shock with $M_s=3$ and $\beta_{p0}=16$ the efficiency of electron irreversible heating is nearly insensitive to the mass ratio, up to $m_i/m_e=1600$. For $m_i/m_e=49$, we choose the transverse size of the box $L_y$ to be $\sim 150 \, c/\omega_{pe}\sim 21\, c/\omega_{pi}$, which is sufficient to capture the growth of proton instabilities in the downstream.

\begin{figure}
\begin{center}
\includegraphics[width=0.4\textwidth]{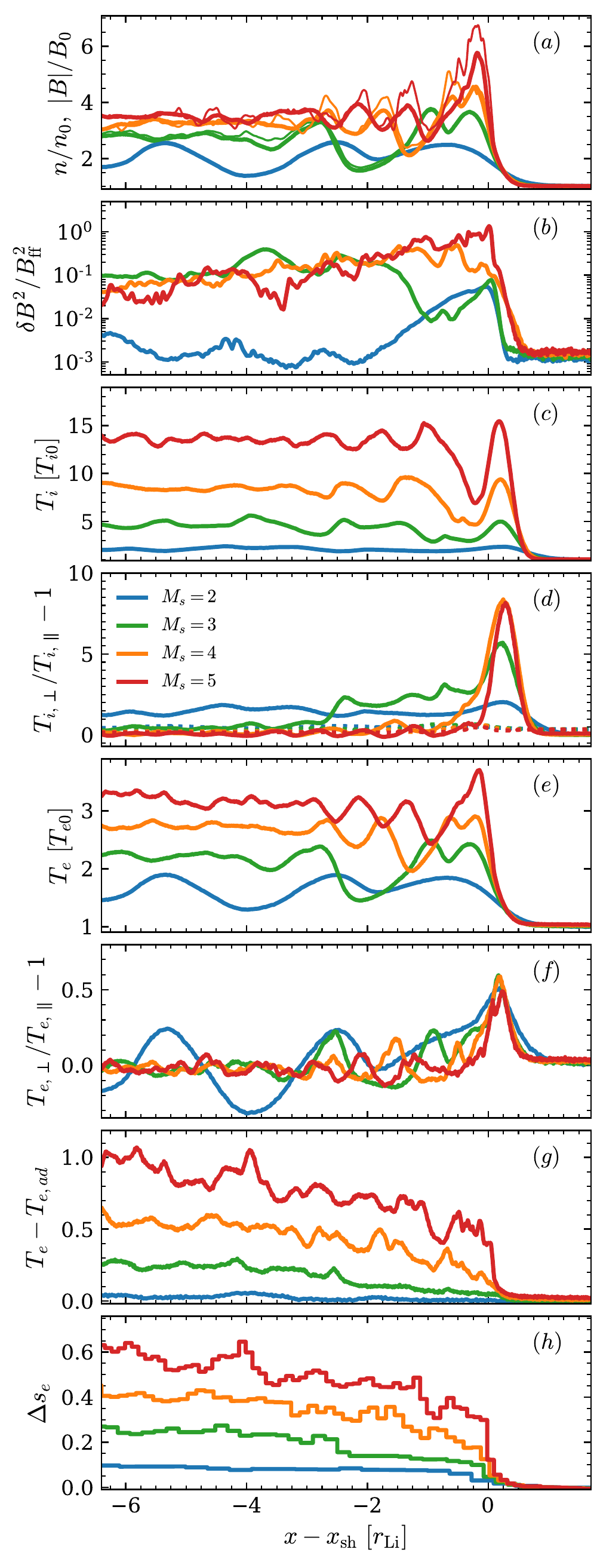}
\end{center}
\caption{
Dependence on $M_s$ of various $y$-averaged quantities, from our shock simulations $\mathtt{Ms2beta16}$, $\mathtt{Ms3beta16}$, $\mathtt{Ms4beta16}$ and $\mathtt{Ms5beta16}$,  at $t=22\,\Omega_{ci}^{-1}$ (the legend is in panel (d)). The $x$ coordinate (aligned with the shock direction of propagation) is measured relative to the shock location $x_{\rm sh}$, in units of the proton Larmor radius $\rli$. From top to bottom, we plot: (a) number density (thick lines) and magnetic field strength (thin lines); (b) energy in magnetic fluctuations, normalized to the energy of the frozen-in  field; (c) mean proton temperature; (d) proton temperature anisotropy (with dotted lines representing the marginal stability threshold in \eq{upperbound}); (e) mean electron temperature; (f) electron temperature anisotropy; (g) excess of electron temperature beyond the adiabatic prediction for an isotropic gas; (h) change in electron entropy. The efficiency of electron irreversible heating increases monotonically with $M_s$. 
}\label{fig:shock_Mss}
\end{figure}

\begin{figure}
\begin{center}
\includegraphics[width=0.5\textwidth]{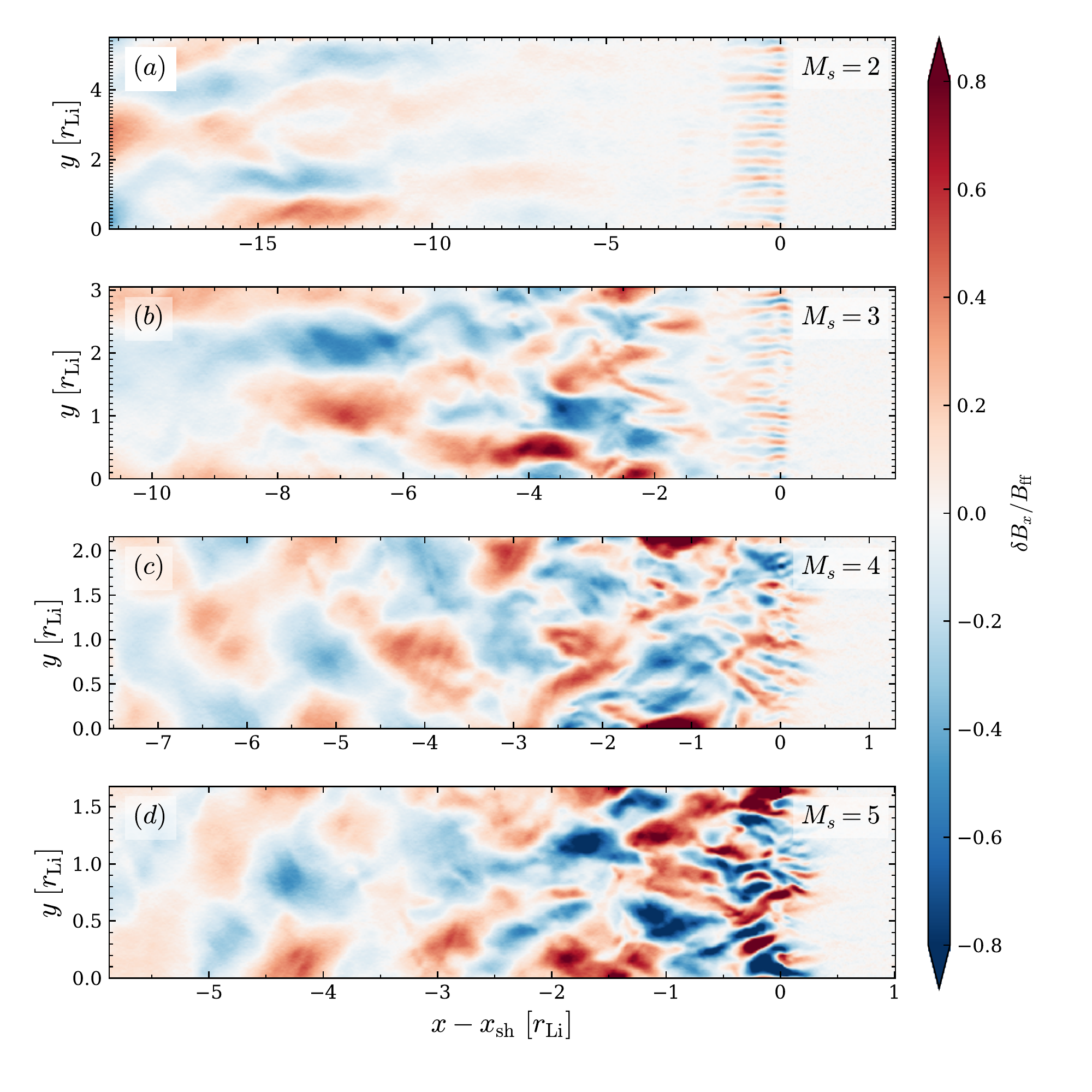}
\end{center}
\caption{
Dependence on $M_s$ of 
the 2D structure of magnetic field fluctuations $\delta B_x/B_{\rm ff}$ in the shock simulations 
$\mathtt{Ms2beta16}$, $\mathtt{Ms3beta16}$, $\mathtt{Ms4beta16}$, $\mathtt{Ms5beta16}$
at $t  = 22\, \Omega_{ci}^{-1}$. 
The $x$ coordinate is measured relative to the shock location $x_{\rm sh}$; both $x$ and $y$ coordinates are normalized to the proton Larmor radius $r_{\rm Li}$.  Notice that the $x$ and $y$ extents of the box are different for different $M_s$. }\label{fig:Msfields}
\end{figure}

\subsection{Dependence on $M_s$}\label{sec:Mstrend}
In this subsection, we compare the results of shock simulations with fixed $\beta_{p0}=16$
and varying Mach number from $M_s=2$ up to $5$ (runs $\mathtt{Ms2beta16}$, $\mathtt{Ms3beta16}$, $\mathtt{Ms4beta16}$ and $\mathtt{Ms5beta16}$ in Table \ref{table:shockruns}). We employ a reduced mass ratio of $m_i/m_e=49$, but in \app{mime} we show that identical results are obtained for a higher value of the mass ratio, $m_i/m_e=200$. 

Figure \ref{fig:shock_Mss} shows the $y$-averaged profiles of various quantities in the shock at time $\Omega_{ci}t=22$, as a function of the $x$ coordinate relative to the shock location $x_{\rm sh}$, in units of the proton Larmor radius $r_{\rm Li}$ defined in \eq{rLi}. Panel (a) shows the profiles of density (thick lines, see the legend in panel (d)) and magnetic field strength (thin lines with the same color coding as the thick lines). In agreement with the Rankine-Hugoniot relations, the density jump is larger for higher $M_s$. As a result of flux freezing alone, one would expect that the magnetic field be $B_{\rm ff}=(n/n_0) B_0$, i.e., its spatial profile should be identical to the density profile. The fact that the field strength $|B|$ in \fig{shock_Mss}(a) exceeds the expectation from flux freezing is to be attributed to magnetic fluctuations. Since the deviation is more pronouced at high $M_s$, whereas thin and thick lines overlap at low $M_s$, we expect from panel (a) that the strength of magnetic field fluctuations should increase with Mach number. 

This is confirmed in \fig{Msfields}, where we present the 2D plots of $\delta B_x/B_{\rm ff}$ for the same simulations as in \fig{shock_Mss}. It is apparent that the strength of long-wavelength fluctuations steadily increases with Mach number (i.e., from top to bottom). Such waves --- a combination of proton cyclotron modes and mirror modes --- accompany the relaxation of the proton temperature anisotropy. As we have discussed in Section \ref{sec:setupmodel}, protons are expected to be highly anisotropic in the immediate post-shock region, with a degree of anisotropy that scales as $T_{\perpi}/T_\pari\propto M_s^2$. The fact that shocks with higher $M_s$ lead to stronger proton anisotropy has two consequences: (\textit{i}) the larger amount of free energy in proton anisotropy can generate stronger proton waves, as indeed observed in \fig{Msfields} and \fig{shock_Mss}(a); (\textit{ii}) as predicted by linear theory (see \app{ICI}), the waves grow faster for higher levels of anisotropy (so, higher $M_s$). In fact, \fig{Msfields} shows that the peak of wave activity is located right at the shock for high Mach numbers
($M_s=4$ and 5), and it shifts farther and farther downstream for lower and lower Mach numbers (it lies at $\xshnorm\sim -2.5$ for  $M_s=3$ and at $\xshnorm\lesssim -10$ for  $M_s=2$), due to the slower and slower wave growth. 

For $M_s=2$ and 3, the wave pattern in the shock ramp is dominated by short-wavelength electron whistler waves, rather than by the long-wavelength proton waves appearing for $M_s=4$ and 5. In fact, the peak at $x\sim x_{\rm sh}$ in the green and blue lines of \fig{shock_Mss}(b) reflects the energy in whistler modes. For high Mach numbers, proton-driven modes are so strong that they dominate the wave energy right at the shock (orange and red curves in \fig{shock_Mss}(b)), hiding the presence of whistler waves in \fig{Msfields}(c),(d). Still, electron whistler modes remain active in the shock ramp (their pattern appears more clearly at higher mass ratios, where proton and electron scales are better separated, see \app{mime}). There, they mediate efficient entropy production, as we discuss below.

As we have just described, the downstream proton waves are sourced by the proton anisotropy induced at the shock, which is expected to be stronger at higher $M_s$. This trend is confirmed in the peak anisotropy of \fig{shock_Mss}(d) (compare the curves at $x\sim x_{\rm sh}$), with the exception of the red curve of $M_s=5$. Here, anisotropy-driven proton instabilities are so violent that the proton anisotropy is not even allowed to reach its expected peak. Due to pitch angle scattering by the proton cyclotron and mirror modes, the proton anisotropy is reduced below the marginal stability condition
\be\label{eq:upperbound}
\frac{T_\perpi}{T_\pari}-1\simeq \frac{1.1}{\beta_\pari^{0.55}}
\ee
which is indicated with dotted lines in \fig{shock_Mss}(d). Since proton-driven waves are stronger for higher $M_s$, the relaxation toward the marginal stability threshold is faster for higher $M_s$, in analogy to what we discussed in \sect{waveM}. The case of $M_s=2$, where proton modes are the weakest, is the only one where the degree of proton anisotropy remains significant (see the blue line in \fig{shock_Mss}(d) in the far downstream). 

Given that protons in the far downstream are generally isotropic, their temperature can be properly quantified by the isotropic-equivalent estimate
\be\label{eq:protiso}
T_i=\frac{2\,T_\perpi+T_\pari}{3}
\ee
which we present in \fig{shock_Mss}(c). The trend seen in the plot, i.e., $T_i/T_{i0}$ increasing with $M_s$, is driven by the fact that the temperature jump predicted by the Rankine-Hugoniot relations for the overall fluid is a monotonic function of $M_s$, in combination with the fact that most of the post-shock fluid energy resides in protons (rather than electrons or proton-driven modes). At sufficiently high Mach numbers, we then expect that $T_i/T_{i0}\propto M_s^2$ (i.e., as predicted by the Rankine-Hugoniot relations), a trend confirmed by \fig{shock_Mss}(c). 

So far, we have mostly focused on the proton physics. As regard to electrons, we find that the post-shock electron temperature is a monotonically increasing function of $M_s$ (Figure \ref{fig:shock_Mss}(e)). This might just follow from the dependence on $M_s$ of the adiabatic heating efficiency, since the density compression increases with $M_s$ (Figure \ref{fig:shock_Mss}(a)).

However, we find that the efficiency of irreversible electron heating is also higher at larger $M_s$. In panel (h) we present the electron entropy profile as measured directly from our simulations using \eq{entre}, while in panel (g) we show the excess of electron temperature $T_e=(2\,T_\perpe+T_\pare)/{3}$ beyond the adiabatic prediction appropriate for a 3D non-relativistic gas 
\be
\frac{T_{e,{\rm ad}}}{T_{e0}}=\left(\frac{n_e}{n_{e0}}\right)^{2/3}~.
\ee
The fact that the efficiency of electron irreversible heating increases with Mach number can be promptly understood from the results obtained in \sect{ramp} and \sect{waves}. First of all, the electron fluid suffers a faster compression while passing through the ramp of a high-$M_s$ shock, as compared to a low-$M_s$ shock. As we have shown in \sect{rampM} (case A), this drives the electrons to larger levels of anisotropy, resulting in stronger whistler waves and faster rates of adiabatic breaking, which leads to more efficient entropy production. In addition, the highly anisotropic protons present in high-$M_s$ shocks generate strong proton modes, as we have discussed in  \sect{waveM} (case B). The resulting field amplification provides another channel to induce electron anisotropy and ultimately leads to electron entropy increase. The stronger proton-driven modes at higher $M_s$ result in higher efficiencies of electron irreversible heating (\sect{waveM}).

While the first mechanism (case A) is present for all the values of Mach number that we investigate, the second (case B) does not operate for $M_s=2$. Here, the post-shock proton anisotropy is weak, and the strength of the resulting proton modes is insufficient to drive the electrons above the threshold of the whistler instability. In the absence of pitch angle scattering to break the adiabatic invariance, the electron entropy (blue line in \fig{shock_Mss}(h)) does not change behind the shock ramp (where entropy production is induced by compression, as in case A). The same had been observed in \sect{waveM}.

In the shock with $M_s=3$ (green line in panel (h)), case A controls the entropy increase in the shock ramp, whereas proton modes (so, case B) are responsible for the additional entropy jump seen at $\xshnorm\sim -2.5$. For $M_s=4$ and 5 (orange and red curves, respectively), proton-driven modes grow fast, and in the shock ramp the field amplification that they induce co-exists with the large-scale compression of the upstream field (i.e., case A and B are spatially coincident). This explains why most of the entropy increase for $M_s=4$ and 5 is localized in the shock transition region. However, since the strength of proton modes remains significant for several proton Larmor radii behind the shock, the proton waves 
can occasionally excite local patches of electron anisotropy that exceed the whistler threshold. The resulting whistler activity  can further increase the electron entropy downstream from the shock, in analogy to what we have discussed in \sect{waveM}.

\subsection{Dependence on $\beta_{p0}$}\label{sec:betatrend}
In this subsection we investigate the dependence of our results on plasma beta, at fixed Mach number. We vary $\beta_{p0}$ from 4 up to 32, for two different values of Mach number: $M_s=3$ in Section \ref{sec:Ms3betatrend} and $M_s=5$ in Section \ref{sec:Ms5betatrend}. We demonstrate that both choices of $M_s$ lead to similar conclusions.

\begin{figure}
\begin{center}
\includegraphics[width=0.4\textwidth]{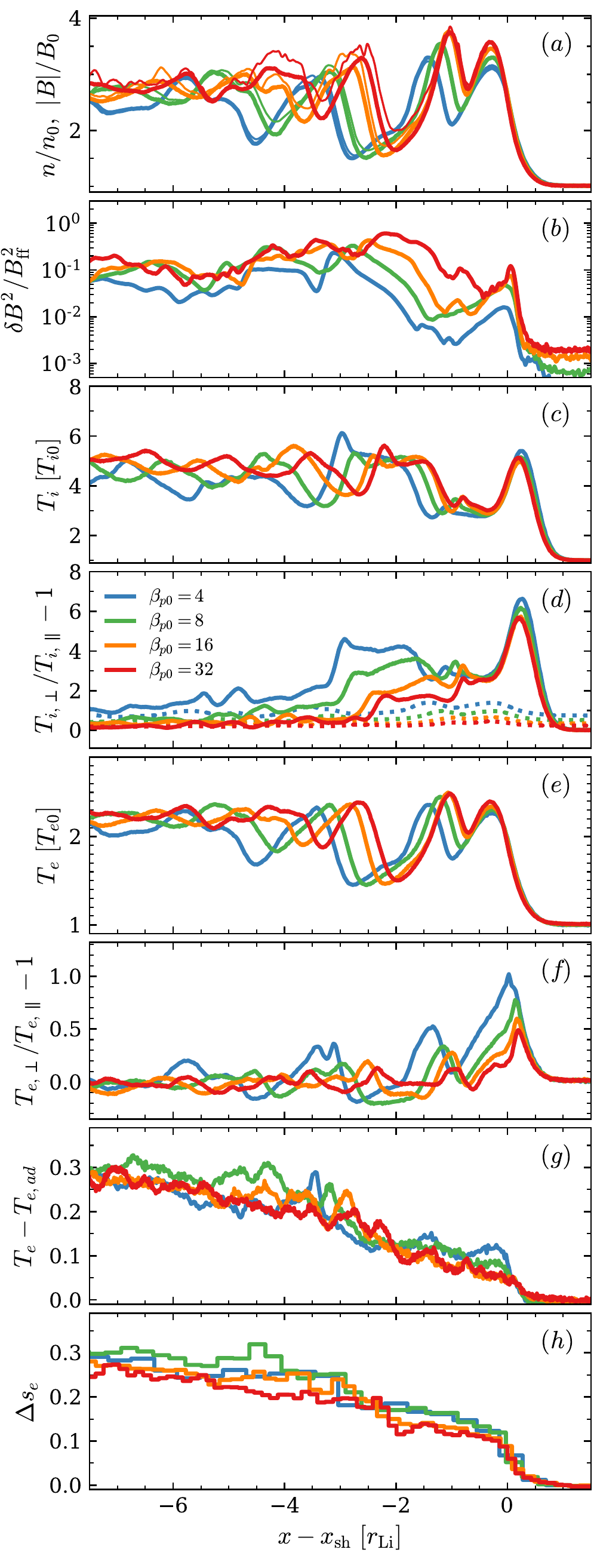}
\end{center}
\caption{
Dependence on $\beta_{p0}$ of various $y$-averaged quantities in our shock simulations with $M_s=3$ (runs $\mathtt{Ms3beta4}$, $\mathtt{Ms3beta8}$, $\mathtt{Ms3beta16}$ and $\mathtt{Ms3beta32}$),  at $t=20.7\,\Omega_{ci}^{-1}$ (the legend is in panel (d)). See the caption of \fig{shock_Mss} for details. Note that the increase in electron entropy is insensitive to the value of $\beta_{p0}$. 
}\label{fig:Ms3betaps}
\end{figure}

\begin{figure}
\begin{center}
\includegraphics[width=0.5\textwidth]{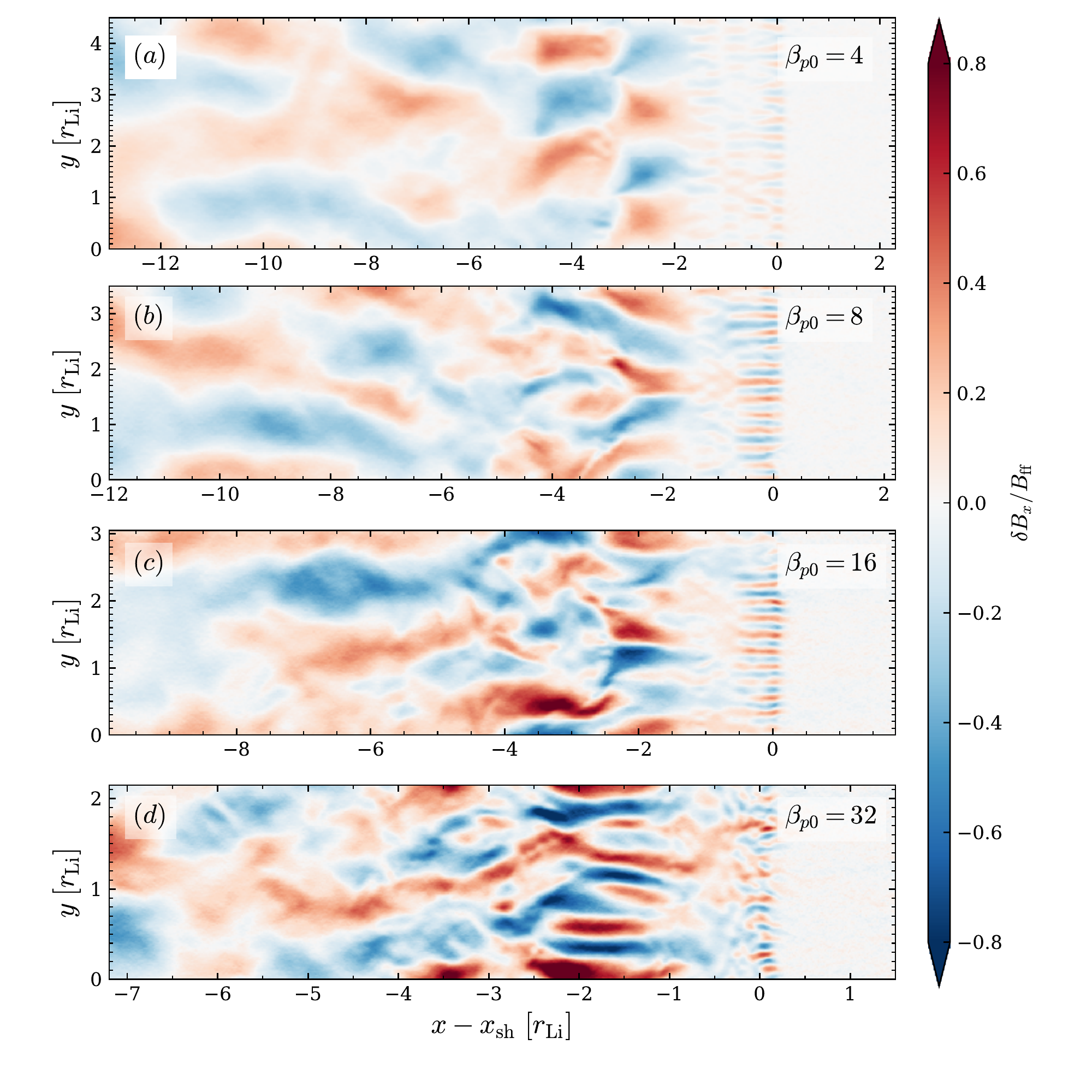}
\end{center}
\caption{
Dependence on $\beta_{p0}$ of 
the 2D structure of magnetic field fluctuations $\delta B_x/B_{\rm ff}$ in the shock simulations 
$\mathtt{Ms3beta4}$, $\mathtt{Ms3beta8}$, $\mathtt{Ms3beta16}$ and $\mathtt{Ms3beta32}$
at $t  = 20.7\, \Omega_{ci}^{-1}$. 
The $x$ coordinate is measured relative to the shock location $x_{\rm sh}$; both $x$ and $y$ coordinates are normalized to the proton Larmor radius $r_{\rm Li}$. Notice that the $x$ and $y$ extents of the box are different for different $\beta_{p0}$. 
}\label{fig:betafields}
\end{figure}

\subsubsection{$M_s=3$}\label{sec:Ms3betatrend}
Figure \ref{fig:Ms3betaps} compares the results of runs $\mathtt{Ms3beta4}$, $\mathtt{Ms3beta8}$, $\mathtt{Ms3beta16}$ and $\mathtt{Ms3beta32}$ having a fixed Mach number $M_s=3$. 
The density compression across the shock is only weakly dependent on $\beta_{p0}$ (Figure \ref{fig:Ms3betaps}(a)), as expected from the Rankine-Hugoniot relations in the limit of high beta. Similarly, the Rankine-Hugoniot jump conditions justify why the post-shock proton temperature (Figure \ref{fig:Ms3betaps}(c)) and the proton anisotropy at the shock (Figure \ref{fig:Ms3betaps}(d) at $x\sim x_{\rm sh}$) are nearly insensitive to $\beta_{p0}$.

Given the relatively weak proton temperature anisotropy attained for $M_s=3$  at the shock ($T_{i,\perp}/T_{i,\parallel}\sim 7$, panel (d)), proton-driven modes grow rather slowly in the downstream, and the magnetic field fluctuations in the shock ramp are powered by the electron whistler instability, for all the values of $\beta_{p0}$ we explore (see \fig{betafields} at the shock; electron whistler modes dominate the peak in magnetic energy seen in \fig{Ms3betaps}(b) at the shock). In analogy to what we discussed in \sect{wavebeta}, the development of proton instabilities is faster at higher $\beta_{p0}$, since the marginal stability threshold is lower (see \eq{upperbound}), and so easier to be exceeded, and because the growth rate is  larger at higher $\beta_{p0}$, for a fixed degree of anisotropy. This explains why the magnetic energy in proton waves peaks closer to the shock at higher $\beta_{p0}$ (in fact, it peaks at $\xshnorm\sim -3$ for the blue line of \fig{Ms3betaps}(b), which refers to $\beta_{p0}=4$, and at $\xshnorm\sim -2$ for the red line,  which refers to $\beta_{p0}=32$). From \fig{Ms3betaps}(b), we also see that proton modes are stronger for higher  $\beta_{p0}$, if normalized to the flux-frozen field. This is just a consequence of the fact that the free energy in proton anisotropy available to source the waves is larger for higher $\beta_{p0}$, when compared to the magnetic energy of the background field (the degree of anisotropy is $\beta_{p0}$-independent, but the proton thermal content obviously increases with $\beta_{p0}$). 

Due to pitch angle scattering by the proton modes, the proton anisotropy drops at a faster rate for higher $\beta_{p0}$ (\fig{Ms3betaps}(d)), since in this case the waves grow faster and are also stronger. In the far downstream, the proton anisotropy settles to the marginal stability threshold of \eq{upperbound}, which is higher for lower $\beta_{p0}$ (see the dotted lines in \fig{Ms3betaps}(d)). It follows that low-$\beta_{p0}$ shocks maintain an appreciable degree of proton anisotropy in the far downstream (see the blue line in \fig{Ms3betaps}(d)). The fact that the resulting adiabatic index will be larger than the value $\Gamma=5/3$ appropriate for a 3D isotropic gas (and so, the plasma will be less compressible) explains why the blue curve in the density profile of panel (a) lies below the other lines. 

The adiabatic heating of electrons will follow the same trend as the density compression of panel (a). As regard to the efficiency of electron irreversible heating, we find that it displays a weak dependence on plasma beta (\fig{Ms3betaps}(g) and (h)). From the findings in \sect{rampbeta} and \sect{wavebeta}, the lack of dependence on $\beta_{p0}$ of the entropy prodution efficiency can be understood as a result of two competing effects. In the case that field amplification is induced by shock-compression of the background field (case A), as appropriate for the shock ramp (for $M_s=3$, proton waves grow farther downstream), higher values of $\beta_{p0}$ generally lead to lower entropy production (\sect{rampbeta}). This is because at higher $\beta_{p0}$ the electron whistler instability can be triggered at lower levels of temperature anisotropy (see the dependence on beta of the stability threshold in \eq{threshwhis}), a trend that is indeed seen at the shock in \fig{Ms3betaps}(f). This leads to weaker whistler waves, less dramatic adiabatic breaking, and a slower rate of entropy production. In fact, in the ramp, shocks with lower $\beta_{p0}$ tend to produce more entropy (see \fig{Ms3betaps}(g) and (h) at $x\sim x_{\rm sh}$).

The trend is opposite when field amplification is induced by proton-driven waves (case B, see \sect{wavebeta}). In this case, higher values of $\beta_{p0}$ lead to stronger proton modes (as observed in  \fig{Ms3betaps}(b)), which perform more work onto the electrons, driving them further (and more often) into the unstable region of the whistler mode. In turn, this leads to more efficient entropy production at higher $\beta_{p0}$. This is the reason why at $\xshnorm\sim -2.5$, i.e., at the peak of the energy in proton waves (see \fig{Ms3betaps}(b)), the electron entropy jump is more pronounced for higher $\beta_{p0}$. It is here that shocks of high $\beta_{p0}$, which were lagging behind in electron entropy production, can catch up with low-$\beta_{p0}$ shocks.
The combination of the two opposite effects lead to an efficiency of electron irreversible heating that is nearly independent of $\beta_{p0}$, for $M_s=3$.

\begin{figure}
\begin{center}
\includegraphics[width=0.4\textwidth]{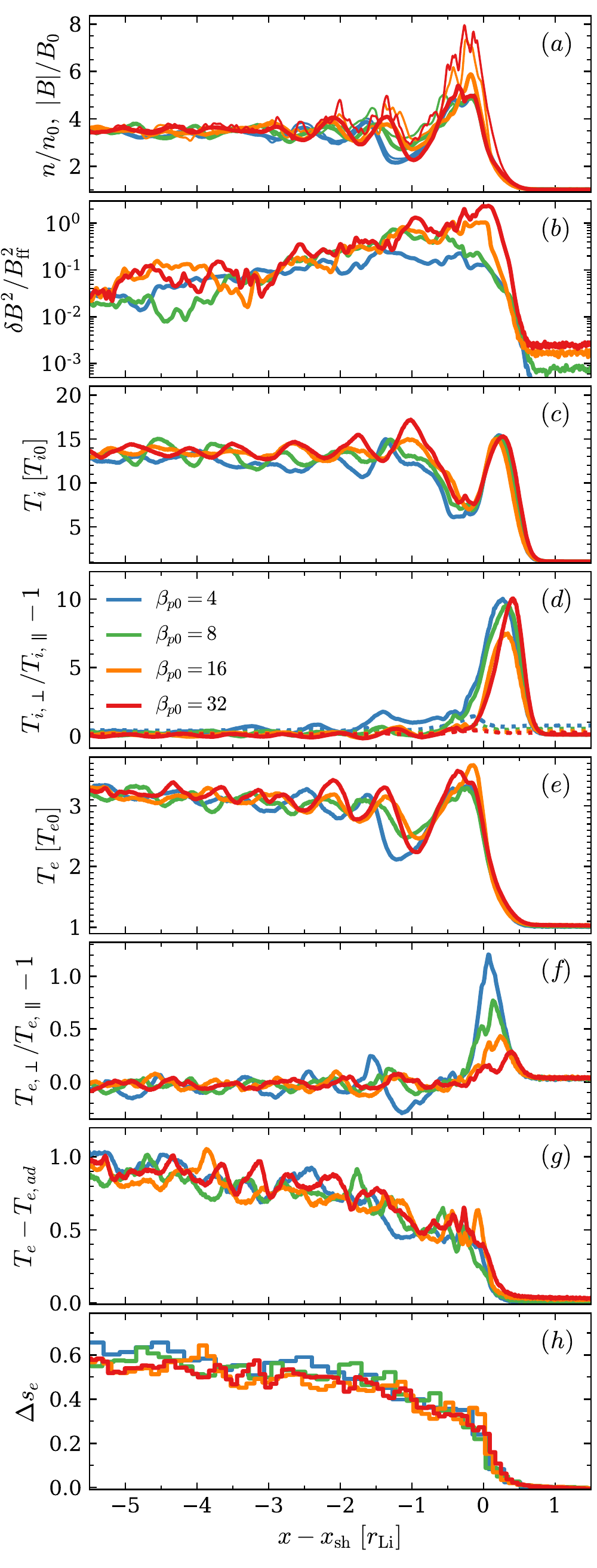}
\end{center}
\caption{
Dependence on $\beta_{p0}$ of various $y$-averaged quantities in our shock simulations with $M_s=5$ (runs $\mathtt{Ms5beta4}$, $\mathtt{Ms5beta8}$, $\mathtt{Ms5beta16}$ and $\mathtt{Ms5beta32}$),  at $t=22\,\Omega_{ci}^{-1}$ (the legend is in panel (d)). See the caption of \fig{shock_Mss} for details. Note that the increase in electron entropy is insensitive to the value of $\beta_{p0}$.}
\label{fig:Ms5betaps}
\end{figure}

\subsubsection{$M_s=5$}\label{sec:Ms5betatrend}
We now demonstrate that the same conclusion --- i.e., the fact that the electron entropy production is independent of $\beta_{p0}$ --- also holds  for shocks with $M_s=5$, by showing in Figure \ref{fig:Ms5betaps} the results of runs $\mathtt{Ms5beta4}$, $\mathtt{Ms5beta8}$, $\mathtt{Ms5beta16}$, $\mathtt{Ms5beta32}$.

The main difference of $M_s=5$ shocks, as compared to their $M_s=3$ counterparts, is that the proton anisotropy at the shock is now so large (Figure \ref{fig:Ms5betaps}(d)) that proton waves grow fast, and their energy peaks right at the shock (Figure \ref{fig:Ms5betaps}(b)). This implies that the two competing effects mentioned above --- i.e., the fact that shock compression leads to more entropy production in low-$\beta_{p0}$ cases, whereas proton waves favor high-$\beta_{p0}$ shocks  --- occur in the same spatial region (specifically, the shock ramp). Despite this difference from the $M_s=3$ cases explored above, the electron entropy production in $M_s=5$ runs still shows a negligible  dependence on $\beta_{p0}$ (panels (g) and (h)).

\begin{figure}
\begin{center}
\includegraphics[width=0.4\textwidth]{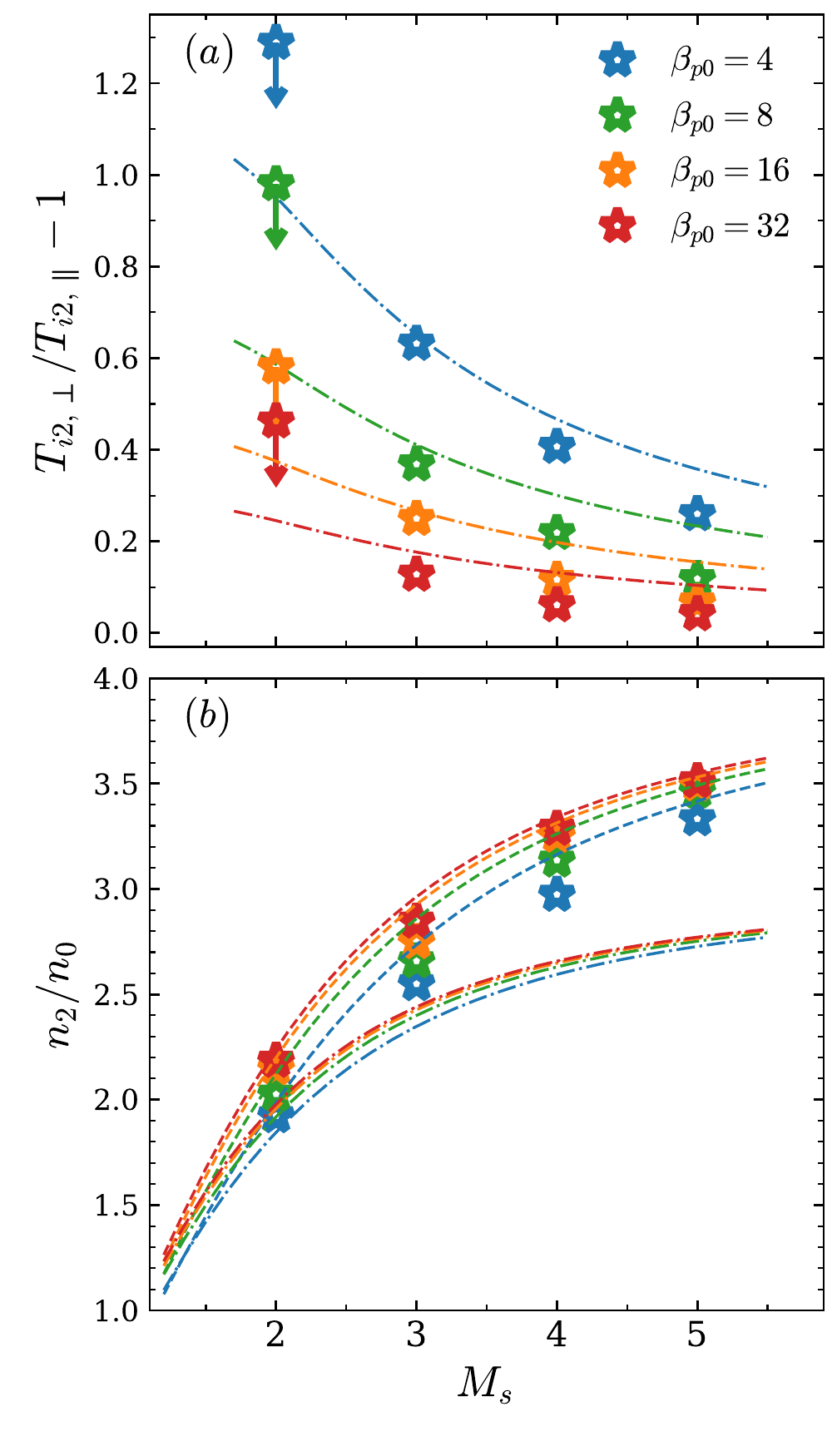}
\end{center}
\caption{Time- and space-averaged downstream proton quantities from the shock simulations with $m_i/m_e=49$ (listed in Table \ref{table:shockruns}), as a function of Mach number $M_s$ (on the horizontal axis) and plasma beta $\beta_{p0}$ (as indicated by the colors, see the legend in the top panel). 
The data points in panel (a) represent the asymptotic proton temperature anisotropy. With dot-dashed lines we plot the predicted upper bound from Equation \eqref{eq:anisobound}.
The data points in panel (b) refer to the density compression. For comparison, we show the Rankine-Hugoniot density jump predictions for a 3D gas ($\Gamma=5/3$, dashed lines) and a 2D gas ($\Gamma=2$, dot-dashed lines). 
}\label{fig:summaryion}
\end{figure}

\begin{table*}
\begin{center}
\begin{tabular}{cccccccc}
\hline 
run name & $M_{s}$ & $\beta_{p0}$ & $T_{i2,\perp}/T_{i2,\parallel}-1$ & $n_2/n_0$ & $T_{e2}/T_{e0}$ & $(T_{e2}-T_{e2,\rm ad})/T_{e0}$ & $T_{e2}/T_{i2}$\tabularnewline
\hline 
\hline 
$\mathtt{Ms2beta4}$ & 2 & 4 & 1.29 & 1.92 & 1.60 & 0.05 & 0.81\tabularnewline
$\mathtt{Ms2beta8}$ & 2 & 8 & 0.98 & 2.03 & 1.70 & 0.10 & 0.80\tabularnewline
$\mathtt{Ms2beta16}$ & 2 & 16 & 0.58 & 2.15 & 1.76 & 0.10 & 0.79\tabularnewline
$\mathtt{Ms2beta32}$ & 2 & 32 & 0.46 & 2.19 & 1.79 & 0.10 & 0.79\tabularnewline
$\mathtt{Ms3beta4}$ & 3 & 4 & 0.63 & 2.55 & 2.18 & 0.31 & 0.49\tabularnewline
$\mathtt{Ms3beta8}$ & 3 & 8 & 0.37 & 2.66 & 2.26 & 0.34 & 0.48\tabularnewline
$\mathtt{Ms3beta16}$ & 3 & 16 & 0.25 & 2.75 & 2.27 & 0.30 & 0.46\tabularnewline
$\mathtt{Ms3beta32}$ & 3 & 32 & 0.13 & 2.85 & 2.27 & 0.27 & 0.46\tabularnewline
$\mathtt{Ms4beta4}$ & 4 & 4 & 0.41 & 2.97 & 2.65 & 0.58 & 0.32\tabularnewline
$\mathtt{Ms4beta8}$ & 4 & 8 & 0.22 & 3.14 & 2.71 & 0.57 & 0.32\tabularnewline
$\mathtt{Ms4beta16}$ & 4 & 16 & 0.12 & 3.25 & 2.72 & 0.52 & 0.32\tabularnewline
$\mathtt{Ms4beta32}$ & 4 & 32 & 0.06 & 3.29 & 2.70 & 0.49 & 0.31\tabularnewline
$\mathtt{Ms5beta4}$ & 5 & 4 & 0.26 & 3.33 & 3.11 & 0.88 & 0.24\tabularnewline
$\mathtt{Ms5beta8}$ & 5 & 8 & 0.12 & 3.47 & 3.13 & 0.84 & 0.23\tabularnewline
$\mathtt{Ms5beta16}$ & 5 & 16 & 0.07 & 3.50 & 3.12 & 0.81 & 0.23\tabularnewline
$\mathtt{Ms5beta32}$ & 5 & 32 & 0.04 & 3.51 & 3.19 & 0.88 & 0.23\tabularnewline
\hline 
\end{tabular}
\caption{Key results of the shock simulations with $m_i/m_e=49$, as discussed in \sect{disc}.}\label{table:summarytab}
\end{center}
\end{table*}

\section{Key Results on Proton Anisotropy and Electron Heating}\label{sec:disc}
\begin{figure}
\begin{center}
\includegraphics[width=0.4\textwidth]{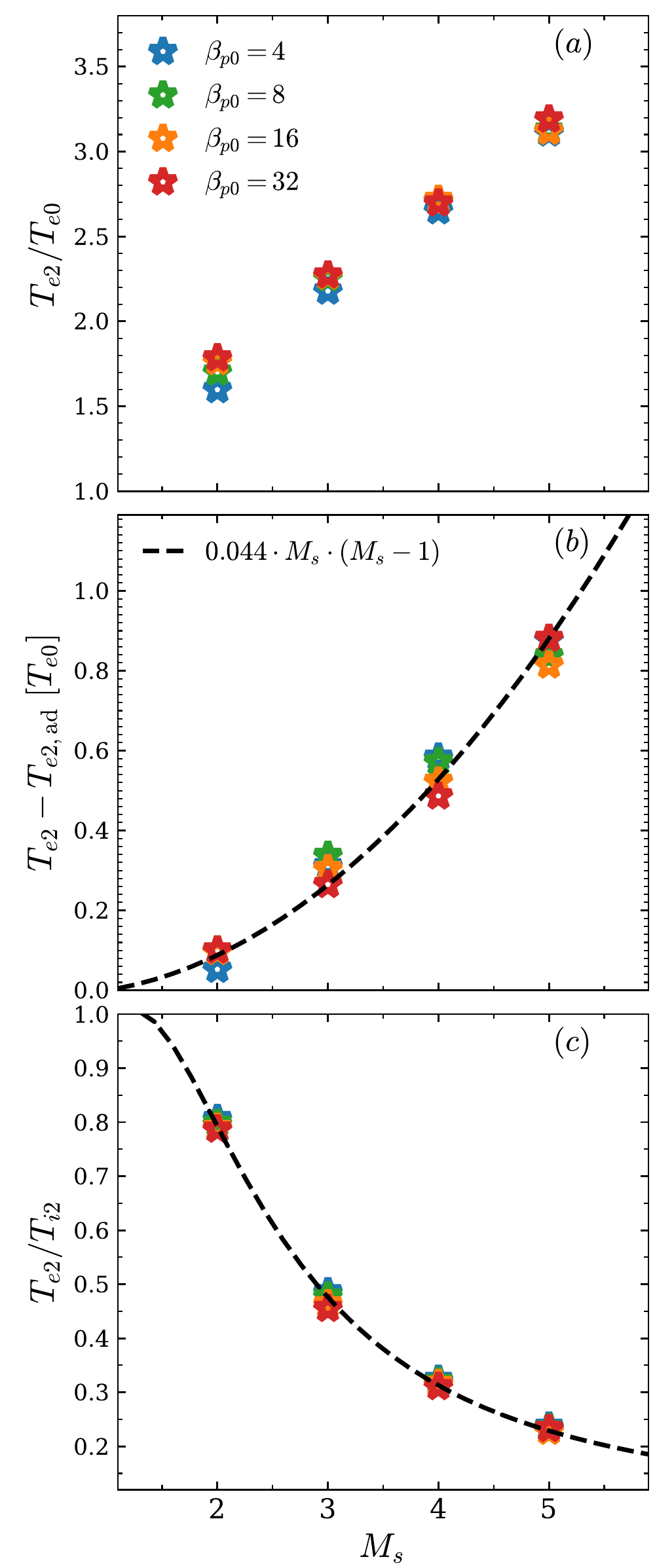}
\end{center}
\caption{Time- and space-averaged downstream electron quantities from the shock simulations with $m_i/m_e=49$ (listed in Table \ref{table:shockruns}), as a function of Mach number $M_s$ (on the horizontal axis) and plasma beta $\beta_{p0}$ (as indicated by the colors, see the legend in the top panel). 
Panel (a) shows the total electron temperature jump across the shock. 
Panel (b) shows the electron temperature excess over the adiabatic expectation of a 3D gas. For comparison, we show  with a dashed black line our proposed fit of Equation \eqref{eq:Tefit}.
Panel (c) shows the post-shock electron-to-proton temperature ratio. For comparison, we show our prediction (Equation \eqref{eq:Tratio}) with a  dashed black line. 
}\label{fig:summaryelec}
\end{figure}

In this section, we summarize our shock results, combining the dependences on $\beta_{p0}$ and $M_s$. We indicate with the subscript ``0'' quantities measured ahead of the shock, and with ``2'' quantities far downstream, computed by spatially averaging in the region where the proton anisotropy has dropped below the threshold in \eq{upperbound}. We then time-average the spatial averages. In the runs with $M_s=2$, where the proton anisotropy is yet to decrease below the threshold in \eq{upperbound} even at times as large as $t=40\,\Omega_{ci}^{-1}$, we compute the quantities with subscript ``2'' as spatial averages in the region at $\xshnorm\lesssim -20$. This particular choice of the averaging region is unlikely to affect our estimates for the efficiency of electron irreversible, since proton waves are too weak to lead to appreciable entropy production in the far downstream; in fact, the only significant increase in electron entropy for  $M_s=2$ occurs at the shock ramp (see the blue line of \fig{shock_Mss}(h)). Our key results are shown in Figures \ref{fig:summaryion} and \fign{summaryelec} and summarized in Table \ref{table:summarytab}.

Figure \ref{fig:summaryion}(a) presents the residual proton temperature anisotropy in the far downstream. As mentioned above, since $M_s=2$ shocks have yet to relax to the threshold in \eq{upperbound}, our measurements are merely upper limits (as indicated by the arrows). For the other cases, we obtain a firm measurement of the proton anisotropy. We find that the residual proton anisotropy decreases with increasing $\beta_{p0}$ (indicated by the colors, see the legend in panel (a)) and increasing $M_s$ (indicated by the horizontal axis). These trends are properly captured by the dot-dashed lines (with the same color coding as the symbols), which are obtained analytically as follows.

In analogy to the discussion in Section \ref{sec:setupmodel}, we assume that in the immediate downstream region the proton plasma can be described as a 2D gas, with perpendicular temperature as in \eq{perpe2} and parallel temperature as in \eq{pare2}.\footnote{The expression in  \eq{perpe2} assumes that electrons are heated adiabatically and that the energy in proton waves is much smaller than the proton thermal content. Both assumptions are reasonably satisfied in the parameter regime we explore.}. We also assume that the proton thermal content stays constant during the relaxation of the proton anisotropy. Even though protons convert part of their energy into magnetic fluctuations, the wave energy in the regime $\beta_{p0}\gg1$ of interest here is much smaller than the proton thermal energy, so our assumption is satisfied. If we indicate with $T_{i2}$ the isotropic-equivalent proton temperature as in \eq{protiso} (which, as we said, stays the same during the  relaxation of proton anisotropy), we can define the plasma beta in the far downstream as 
\be
\beta_{i2}\equiv \frac{8\pi n_2 T_{i2}}{B_2^2}\,,
\ee
where $n_2$ and $B_2$ are the density and field strength in the far downstream. There, proton waves have decayed, and $B_2$ is the flux-frozen field strength, which can be determined via the shock density jump
\be
\frac{B_2}{B_0}=\frac{n_2}{n_0}=r_{\rm RH, \Gamma=5/3}~~,
\ee 
where $\Gamma=5/3$ is appropriate if the residual proton anisotropy is small in the far downstream, as in most of our cases. It follows that
\be
\beta_{i2}=\frac{T_{i2}}{T_0}\frac{\beta_{p0}}{2}\frac{1}{r_{\rm RH, \Gamma=5/3}}~~,
\ee
and the threshold in \eq{upperbound}, which we rewrite as
\be\label{eq:anisobound}
\frac{T_{i,\perp}}{T_{i,\parallel}} - 1 - \frac{1.1}{\left[3\beta_{i2}/(2\,T_{i,\perp}/T_{i,\parallel}+1)\right]^{0.55}} \simeq 0 ~~,
\ee
may be solved for the anisotropy $T_{i2,\perp}/T_{i2,\parallel}-1$ in the far downstream. The solutions of Equation \eqref{eq:anisobound} at different $\beta_{p0}$ as a function of $M_s$ are plotted as dot-dashed lines in Figure \ref{fig:summaryion}(a). They capture the trend of decreasing asymptotic proton temperature anisotropy with increasing $\beta_{p0}$ and $M_s$. Since \eq{upperbound}
 (or equivalently, \eq{anisobound}) sets an upper limit for the asymptotic proton anisotropy, it is also expected that our data points (apart from the $M_s=2$ runs where the proton anisotropy is yet to relax) should lie below the upper bound prescribed by \eq{upperbound}.

Since the post-shock protons are not perfectly isotropic (Figure \ref{fig:summaryion}(a)), one might expect that the density jump $n_2/n_0$ from the upstream to the far downstream  (Figure \ref{fig:summaryion}(b)) will lie in between the predictions for a $\Gamma = 2$ gas (anisotropic, with two degrees of freedom) and  for a $\Gamma = 5/3$ gas (perfecly isotropic, with three degrees of freedom). In panel (b), we plot the former with dot-dashed lines and the latter with dashed curves. Indeed, our data points lie in between the two sets of curves. Since the post-shock protons tend to be more isotropic for higher $\beta_{p0}$ and $M_s$ (panel (a)), our data points in panel (b) tend to move closer to the $\Gamma = 5/3$ curves for either larger  $\beta_{p0}$ (compare the blue points for $\beta_{p0}=4$ with the red points for $\beta_{p0}=32$) or higher $M_s$ (e.g., compare $M_s=3$ with $M_s=5$). The overall increase in $n_2/n_0$ with Mach number is simply a consequence of the Rankine-Hugoniot relations.

We next summarize in Figure \ref{fig:summaryelec} the dependence on $\beta_{p0}$ and $M_s$ of the electron heating efficiency (the legend is in panel (a)). The overall electron temperature jump $T_{e2}/T_{e0}$, which includes both adiabatic and irreversible contributions, shows a weak dependence on $\beta_{p0}$ and a systematic increase with $M_s$. In the regime of low Mach numbers investigated in this work, most of the electron temperature increase is contributed by adiabatic heating (compare the overall temperature jump in panel (a) with the irreversible contribution in panel (b)), and so the temperature jump should be nearly equal to
\be
T_{e2,\rm ad}/T_{e0} = (n_2/n_0)^{2/3}=r_{\rm RH, \Gamma=5/3}^{2/3}
\ee
as a result of density compression in a 3D gas. In fact, the mild increase in $T_{e2}/T_{e0}$ with $\beta_{p0}$ at fixed $M_s$ is primarily a consequence of the dependence of $n_2/n_0$ on plasma beta (see \fig{summaryion}(b)).

At high values of $M_s$, the contribution of adiabatic heating will saturate, since $n_2/n_0\rightarrow 4$ for a 3D gas. Here, most of the electron heating will be controlled by irreversible processes (i.e., associated with entropy increase). Panel (b) shows the efficiency of electron irreversible heating, quantified by the electron temperature increase beyond the adiabatic expectation. As discussed in \sect{shock}, the efficiency of electron irreversible heating increases with Mach number and is nearly insensitive to plasma beta, in the regime explored in this work (as a reminder, the fact that it is nearly independent of $\beta_{p0}$ is due to the opposite effects of shock-compression and proton waves, which tend to cancel out). Combining all the data, we
find that the efficiency of electron irreversible heating can be fitted quite well with the following simple function,
\begin{equation}\label{eq:Tefit}
\frac{ T_{e2} -T_{e2,\rm ad} }{T_{e0}} \simeq  0.044\,M_s\, (M_s -1)\equiv \Delta t_{e2,\rm irr}~~,
\end{equation}
with no appreciable dependence on $\beta_{p0}$. The above fitting function is shown in panel (b) with a dashed black line.
Note that this fitting formula exhibits the correct behavior in the limit $M_s\rightarrow1$: in the absence of a shock, the efficiency tends to zero.

Given our empirical fit in \eq{Tefit}, the overall electron temperature jump in the shock is given by
\begin{equation}\label{eq:Tetot}
\frac{T_{e2}}{T_{e0}} = r_{{\rm RH},\Gamma=5/3}^{2/3} + \Delta t_{e2,\rm irr}~~,
\end{equation}
and the resulting ratio of electron and proton temperatures in the far downstream will be
\begin{equation}\label{eq:Tratio}
\frac{T_{e2}}{T_{i2}}  = \frac{r_{{\rm RH},\Gamma=5/3}^{2/3} + \Delta t_{e2,\rm irr}}{2 \,\Delta t_{{\rm RH},\Gamma=5/3} -r_{{\rm RH},\Gamma=5/3}^{2/3} - \Delta t_{e2,\rm irr}}~~,
\end{equation}
where the proton temperature is obtained by subtracting the electron contribution from the Rankine-Hugoniot temperature jump $\Delta t_{{\rm RH},\Gamma=5/3}$ of the overall fluid (assuming a 3D gas). This prediction is plotted in panel (c) with a black dashed line (we assume $\beta_{p0}=32$ in calculating the Rankine-Hugoniot  jump conditions, but the curve will be nearly the same as long as $\beta_{p0}\gg1$). The prediction matches very well with the simulation results.

We conclude with two important comments. First, we remark that the results of this section have been obtained for our reference value of the mass ratio, $m_i/m_e=49$. In Paper I, we explicitly demonstrated that the electron entropy increase in our reference shock with $M_s=3$ and $\beta_{p0}=16$ is nearly insensitive to the mass ratio, from $m_i/m_e=49$ up to $m_i/m_e=200$ (see also \app{mime} for the same conclusion in the case of $M_s=5$ shocks with different $\beta_{p0}$). In addition, in Paper I we were able to extrapolate our results up to the realistic mass ratio, in controlled periodic box experiments meant to mimic the two possible scenarios for field amplification (i.e., shock-compression of the upstream field, or field amplification due to proton waves in the downstream). We found that the electron irreversible heating efficiency has only a weak dependence on mass ratio --- less than $\sim30\%$ decrease as we increase the mass ratio from $m_i/m_e=49$ up to $m_i/m_e=1600$. Based on this result, we argue that the coefficient in \eq{Tefit} should be reduced by the same fraction $\sim30\%$ for realistic mass ratios, so it will be $\simeq 0.03$ instead of $\simeq 0.044$.

Second, we remind the reader that our results have been obtained for strictly perpendicular shocks. 
In quasi-parallel shocks (i.e., where the angle between the pre-shock field and the shock direction of propagation is $\lesssim 45^\circ$), protons are expected to be efficiently reflected back upstream and accelerated via the Fermi process \citep[e.g.,][]{Caprioli2014,Park2015}. In this regime, efficient electron heating (up to equipartition with the protons) was observed for supernova remnant conditions (i.e., at $M_s$ of a few hundreds and $\beta_{p0}\sim 1$), and similar conclusions should apply in the low-$M_s$ high-$\beta_{p0}$ regime investigated here. In contrast, in quasi-perpendicular shocks, (i.e., where the angle between the pre-shock field and the shock direction of propagation is $\gtrsim 45^\circ$), protons are not efficiently injected into the Fermi process. At low $M_s$, electrons can still be efficiently accelerated, as we have shown in  \citet{Guo2014,Guo2014c}. As long as the non-thermal electrons are energetically sub-dominant, we expect that the conclusions presented in this paper as regard to the electron heating efficiency will still apply for quasi-perpendicular field configurations.

\section{Summary}\label{sec:summ}
In this paper, the second of a series, we have quantified with 2D PIC simulations how the efficiency of electron heating and the post-shock electron-to-proton temperature ratio depend on the shock Mach number $M_s$ and the plasma beta $\beta_{p0}$, in the regime relevant for galaxy cluster shocks. In Paper I, we described the physics of electron heating. In analogy to the so-called ``magnetic pumping'' mechanism \citep{Spitzer1953,Berger1958,Borovsky1986}, we found that two basic ingredients are needed for electron irreversible heating: (\textit{i}) the presence of a temperature anisotropy, induced by field amplification coupled to adiabatic invariance; and (\textit{ii}) a mechanism to break the adiabatic invariance itself. We found that the growth of whistler waves --- triggered by the electron temperature anisotropy induced by field amplification --- was responsible for the violation of adiabatic invariance, and efficient entropy production. 

While Paper I focused only on a reference shock with $M_s=3$ and $\beta_{p0}=16$, here we have extended our study to a wide range of plasma beta ($4\lesssim \beta_{p0}\lesssim 32$) and sonic Mach number ($2\lesssim M_s\lesssim 5$). We first employed periodic box experiments to reproduce, under controlled conditions, the two mechanisms that can drive field amplification in shocks: (\textit{i}) shock-compression of the upstream field, and (\textit{ii}) the growth of proton cyclotron modes accompanying the relaxation of proton temperature anisotropy. Armed with a detailed understanding of the electron heating efficiency in these two scenarios, and of its dependence on $M_s$ and $\beta_{p0}$, we then studied the efficiency of electron entropy production in full shock simulations, where the two mechanisms generally co-exist. 

Our main results are summarized in \sect{disc}. Most importantly, we find that the electron irreversible heating efficiency in shocks is nearly independent of $\beta_{p0}$, and its dependence on $M_s$ can be cast in a simple form: for our reference mass ratio $m_i/m_e=49$, the post-shock electron temperature $T_{e2}$ exceeds the adiabatic expectation $T_{e2,\rm ad}$ by an amount that scales with Mach number as $(T_{e2}-T_{e2,\rm ad})/T_{e0}\simeq 0.044 \,M_s (M_s-1)$, where $T_{e0}$ is the pre-shock temperature (see \eq{Tefit}). As discussed in \sect{disc}, the coefficient should be reduced by $\sim30\%$ when extrapolating to realistic mass ratios (so, it will be $\simeq 0.03$ instead of $\simeq 0.044$).
This can be used to predict the electron-to-proton temperature ratio in the shock downstream (see \eq{Tratio}), with important implications for current and future measurements of electron-proton equilibration in galaxy cluster shocks. 

Although we have only focused on perpendicular field geometries, we argue that our conclusions will also apply for quasi-perpendicular shocks (see \sect{disc}), as long as the non-thermal electrons that are accelerated in such configurations stay energetically sub-dominant. For quasi-parallel geometries, protons are efficiently injected into the Fermi process, and the two species might be led to thermal equilibrium, as found by \citet{Park2015}.
 The robustness of our electron heating mechanism --- and the resulting efficiency of electron irreversible heating --- in supernova remnant shocks ($M_s$ of a few hundreds and $\beta_{p0}\sim1 $) and heliospheric shocks (low Mach number and $\beta_{p0}\sim 1$) remains to be explored.

\acknowledgements
This work is supported in part by the Black Hole Initiative at Harvard University through a grant from the Templeton Foundation.  XG and RN acknowledge support from NASA TCAN NNX14AB47G and NSF grant AST 1312651. LS acknowledges
support from DoE DE-SC0016542, NASA
Fermi NNX-16AR75G, NASA ATP NNX-17AG21G, NSF ACI-1657507, and NSF AST-1716567. The simulations were performed on Habanero at Columbia, the BHI cluster at the Black Hole Initiative, NASA High-End Computing (HEC) Program through the NASA Advanced Supercomputing (NAS) Division at Ames Research Center, and NSF XSEDE resources (grant TG-AST080026N). 

\appendix


\begin{figure}
\begin{center}
\includegraphics[width=0.4\textwidth]{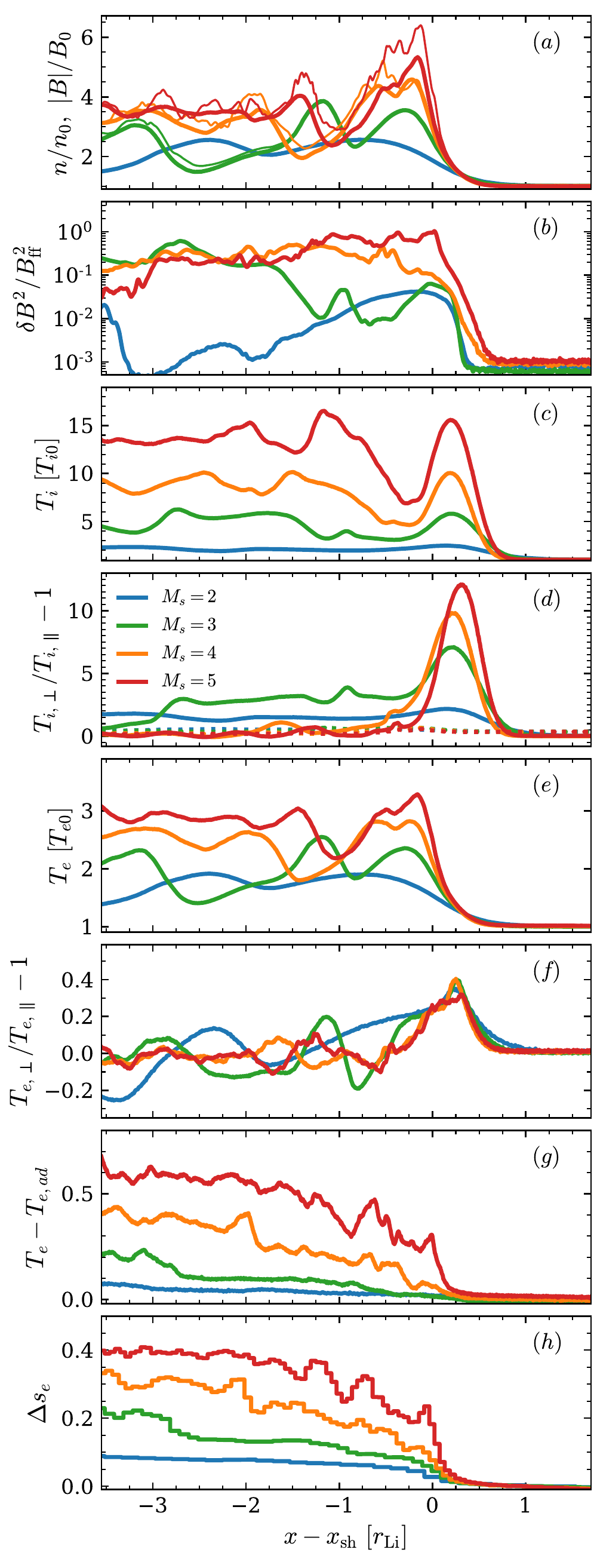}
\end{center}
\caption{
Dependence on $M_s$ of various $y$-averaged quantities, from our shock simulations with $m_i/m_e = 200$ ($\mathtt{mi200Ms2}$, $\mathtt{mi200Ms3}$, $\mathtt{mi200Ms4}$ and $\mathtt{mi200Ms5}$),  at $t=12.9\,\Omega_{ci}^{-1}$  (the legend is in panel (d)). The $x$ coordinate (aligned with the shock direction of propagation) is measured relative to the shock location $x_{\rm sh}$, in units of the proton Larmor radius $\rli$. From top to bottom, we plot: (a) number density (thick lines) and magnetic field strength (thin lines); (b) energy in magnetic fluctuations, normalized to the energy of the frozen-in  field; (c) mean proton temperature; (d) proton temperature anisotropy (with dotted lines representing the marginal stability threshold in \eq{upperbound}); (e) mean electron temperature; (f) electron temperature anisotropy; (g) excess of electron temperature beyond the adiabatic prediction for an isotropic gas; (h) change in electron entropy. If we compare to Figure \ref{fig:shock_Mss}, which employed a lower mass ratio ($m_i/m_e=49$),  we confirm that the shock physics (and in particular, the efficiency of electron irreversible heating) is nearly insensitive to the choice of mass ratio.
}\label{fig:shock_Mss_mi200}
\end{figure}

\begin{figure}
\begin{center}
\includegraphics[width=0.5\textwidth]{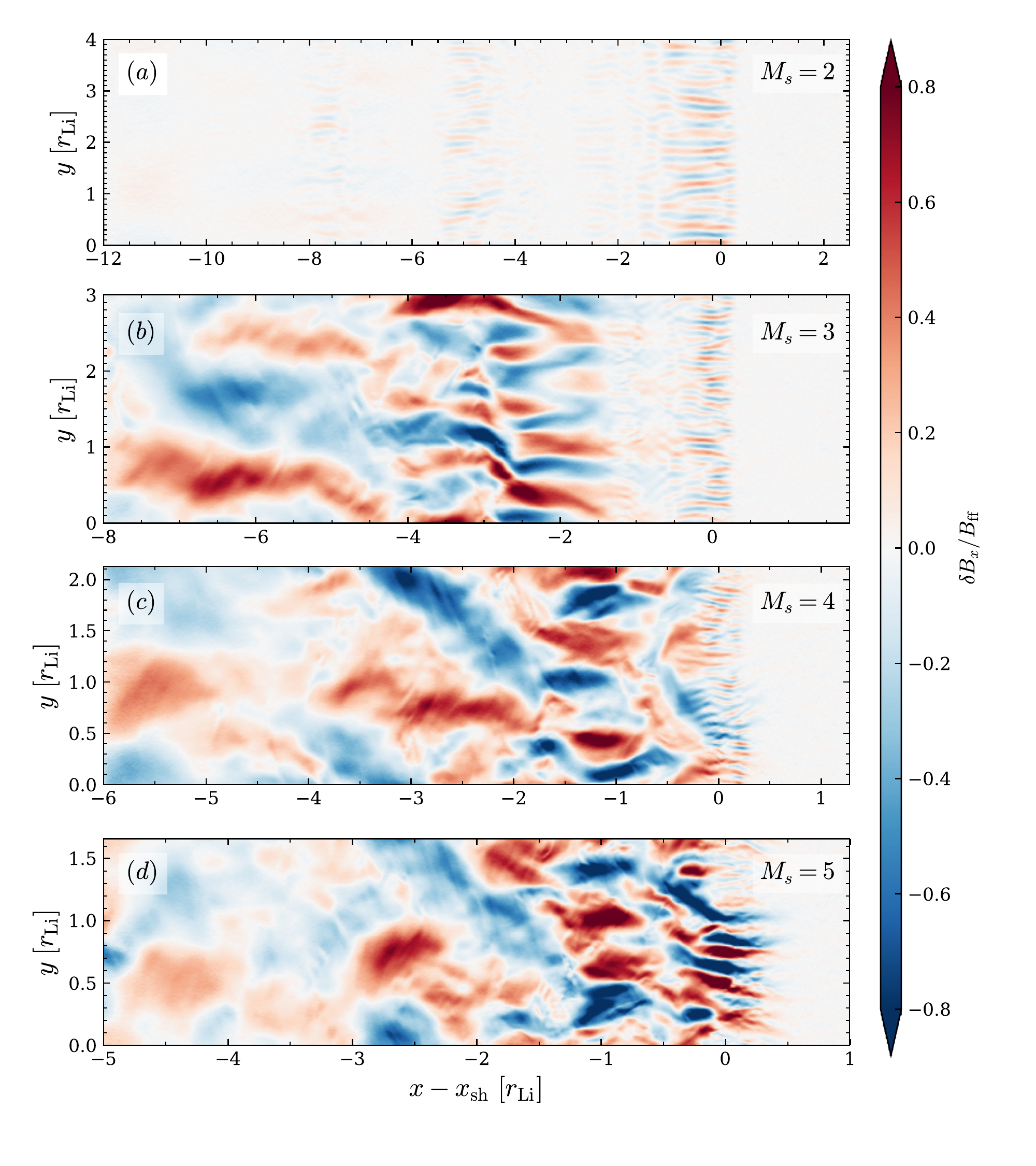}
\end{center}
\caption{For $m_i/m_e=200$, we present the dependence on $M_s$ of 
the 2D structure of magnetic field fluctuations $\delta B_x/B_{\rm ff}$ in the shock simulations 
$\mathtt{mi200Ms2}$, $\mathtt{mi200Ms3}$, $\mathtt{mi200Ms4}$, $\mathtt{mi200Ms5}$
at $t  = 12.9\, \Omega_{ci}^{-1}$. 
The $x$ coordinate is measured relative to the shock location $x_{\rm sh}$; both $x$ and $y$ coordinates are normalized to the proton Larmor radius $r_{\rm Li}$.  Notice that the $x$ and $y$ extents of the box are different for different $M_s$. As compared to Figure \ref{fig:Msfields}, which employed a lower mass ratio ($m_i/m_e=49$),  electron modes in the shock ramp now appear more clearly, due to the larger separation between electron and proton scales.
}\label{fig:Msfields_mi200}
\end{figure}

\section{Dependence on the Mass Ratio}\label{sec:mime}
In this appendix, we investigate the dependence on $M_s$ in shock simulations having $\beta_{p0}=16$ and a higher value of the mass ratio ($m_i/m_e=200$) as compared to the choice $m_i/m_e=49$ employed in the main body of the paper (the runs presented in this appendix are $\mathtt{mi200Ms2}$, $\mathtt{mi200Ms3}$, $\mathtt{mi200Ms4}$ and $\mathtt{mi200Ms5}$ in Table \ref{table:shockruns}).

In Figure \ref{fig:shock_Mss_mi200}, we present the $y$-averaged profiles of various quantities  at $t=12.9\,\Omega_{ci}^{-1}$. Comparing with Figure \ref{fig:shock_Mss} (which employed $m_i/m_e=49$), we see that the profiles are almost identical, for both protons and electrons. This proves that the electron heating physics is insensitive to the mass ratio, as long as proton and electron scales are sufficiently separated (see Paper I for further details on the dependence on mass ratio).

One important advantage of simulations with higher mass ratio is the fact that electron whistler modes in the shock ramp  appear more clearly for $m_i/m_e=200$ (as compared to the case $m_i/m_e=49$), due to the larger separation between electron and proton scales. This is particularly critical at high $M_s$, since proton waves grow right at the shock, and their wavelength is quite small (potentially, approaching electron scales), due to the strong proton anisotropy (see \app{ICI}).
For instance, electron whistler waves are much more apparent in \fig{Msfields_mi200}(c) (having $m_i/m_e=200$) than in \fig{Msfields}(c) (which employed $m_i/m_e=49$).

\section{Linear properties of the electron whistler instability}\label{sec:ewlinear}
According to our electron heating model, the presence of a mechanism to break the electron adiabatic invariance is essential for generating electron entropy. The electron whistler instability is usually invoked to serve this purpose in the shock downstream. In this appendix, we study the linear properties of the whistler instability that are useful for interpreting the simulation results presented in the main body of the paper.

Following \cite{Gary1985}, we solve the dispersion relation for the electron whistler  instability
\begin{eqnarray}
0 &=& D^{\pm } (k, \Omega)\nonumber \\
 &=&\Omega^2 - c^2 k^2 + \omega_{pi}^2 \frac{\Omega}{k v_i}Z(\zeta_i^{\pm})
 + \omega_{pe}^2\frac{\Omega}{k v_{e,\parallel}^2}Z(\zeta_e^{\pm})\nonumber \\
& & + \omega_{pe}^2\left( \frac{T_{e\perp}}{T_{e,\parallel}}-1\right)
\left[1+\zeta_e^{\pm } Z(\zeta_e^{\pm})\right]\,,
\label{eq:ewdispersion}
\end{eqnarray}
where 
$\Omega = \omega + i \gamma$ is the frequency of the instability, 
$k$ is the wavevector, 
$
\zeta_e^{\pm} = (\Omega \pm \Omega_{ce})/k v_{e,\parallel},
$
$\Omega_{ce}=(m_i/m_e)\, \Omega_{ci}$,
$
v_{e,\parallel} = (2\,k_{\rm B}T_{e,\parallel}/m_e)^{1/2},
$
$
\zeta_i^{\pm} = (\Omega \pm \Omega_{ci})/k v_{i},
$
$
v_{i} = (2\,k_{\rm B}T_{i}/m_i)^{1/2}, 
$
and
$Z(\zeta)$ is the plasma dispersion function 
\begin{equation}\label{eq:plasmaZeta}
Z(\zeta) = \frac{1}{\sqrt{\pi}}\int_{-\infty}^{\infty} dx \frac{\exp(-x^2)}{x-\zeta}~.
\end{equation}

We explore how the dispersion relation of the electron whistler instability, i.e., its growth rate $\gamma$ as a function of the wavevetor $k$, varies for different levels of electron temperature anisotropy $T_{e,\perp}/T_{e,\parallel}$, and of electron plasma beta parallel to the magnetic field $\beta_{e,\parallel}$. 
For all calculations, we fix $m_i/m_e =49$ and $v_i = 0.02$ c. 

\begin{figure}
\begin{center}
\includegraphics[width = 0.45\textwidth]{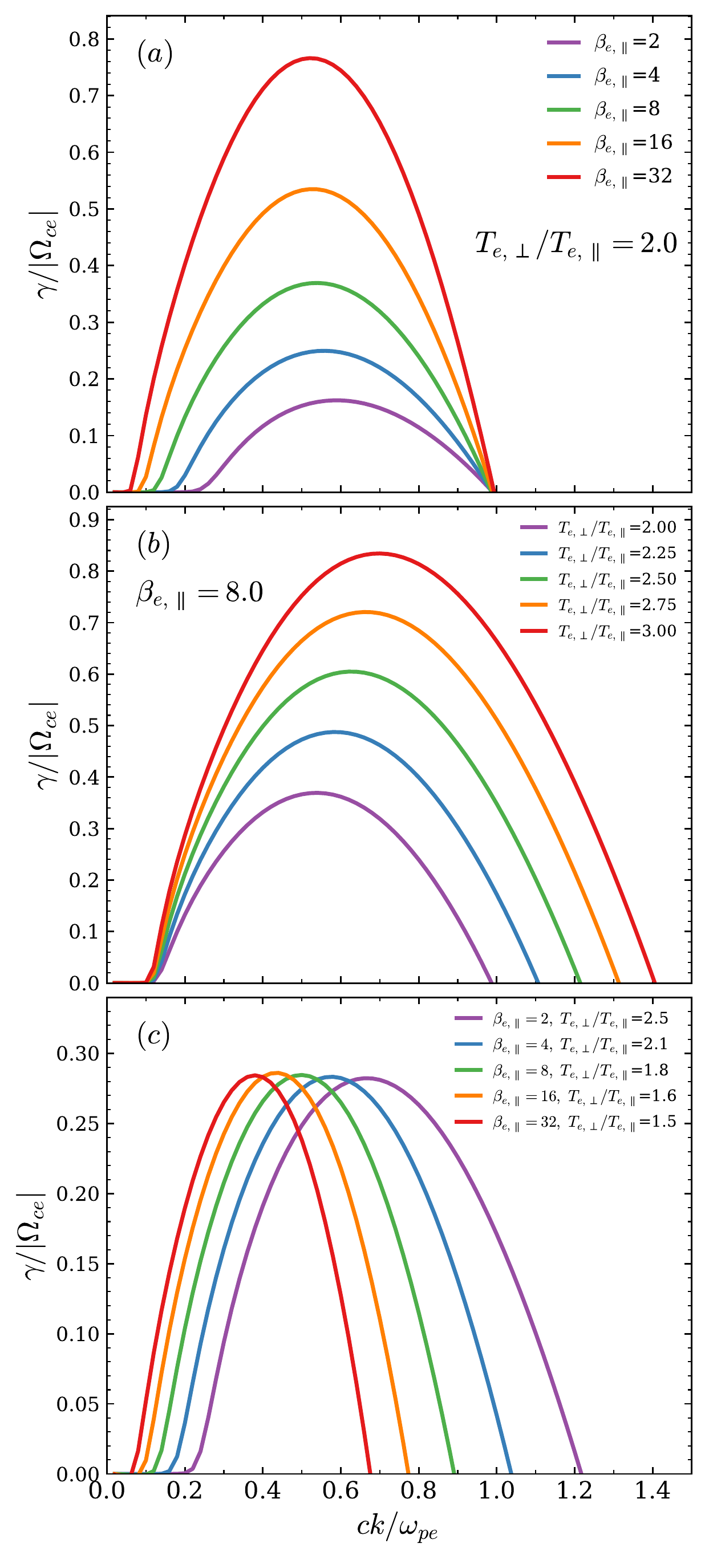}
\caption{Dispersion relation of the electron whistler instability, i.e., the solution of Equation \eqref{eq:ewdispersion}. Panel (a) shows the dependence on $\beta_{e,\parallel}$ at fixed electron temperature anisotropy $T_{e,\perp}/T_{e,\parallel}=2$; Panel (b)  explores the dependence on temperature anisotropy at fixed  $\beta_{e,\parallel}$;  Panel (c) shows the dispersion relation for different combinations of $\beta_{e,\parallel}$ and $T_{e,\perp}/T_{e,\parallel}$ that lead to a fixed maximum growth rate of $\gamma_{max}=0.28\, \Omega_{ce}$.}\label{fig:ewdispersion}
\end{center}
\end{figure}

Figure \ref{fig:ewdispersion}(a) compares the dispersion relation at fixed electron temperature anisotropy $T_{e,\perp}/T_{e,\parallel}=2$ but different $\beta_{e,\parallel}$, ranging from $2$ to $32$. For a given temperature anisotropy, the growth rate of the electron whistler instability increases  monotonically  with $\beta_{e,\parallel}$, and the wavelength of the maximally growing mode increases, or equivalently the wavevector of the maximally growing mode decreases. 
Figure \ref{fig:ewdispersion}(b) depicts the trend of the dispersion relation at fixed  $\beta_{e,\parallel}=8$ but for different levels of electron temperature anisotropy. We see that the growth rate of the electron whistler instability increases monotonically with increasing temperature anisotropy $T_{e,\perp}/T_{e,\parallel}$, and the wavelength of  the maximally growing  mode decreases.  
It follows that, in order to attain a given growth rate of the instability, the required temperature anisotropy  is lower for higher $\beta_{e,\parallel}$. In addition, the wavelength of the maximally growing mode is  longer at higher $\beta_{e,\parallel}$. Indeed,  Figure \ref{fig:ewdispersion}(c) shows that in order to reach a maximum growth rate of $\sim 0.28\, \Omega_{ce}$, the required temperature anisotropy decreases from $2.5$ for $\beta_{e,\parallel}=2$ to $1.5$ for $\beta_{e,\parallel} = 32$. The maximally growing wave vector decreases from $0.7\, \omega_{pe}/c$ down to $0.35\, \omega_{pe}/c$. 

\section{Linear  properties of the proton cyclotron instability}\label{sec:ICI}
In our electron heating model, we require the presence of electron anisotropy. Electron anisotropy can be induced by field amplifcation via the proton cyclotron instability, which naturally occurs in the shock downstream, where it is sourced by proton anisotropy. In  this appendix, we study the linear properties of the proton cyclotron instability and its dependence on plasma beta and on the level of proton temperature anisotropy. 

Following \cite{Davidson1975}, we solve the dispersion relation for the proton cyclotron instability (Equation (3) of \citet{Davidson1975})
\begin{eqnarray}
0 &=& D^{\pm } (k, \Omega) \nonumber \\
&=& \Omega^2 - c^2 k^2 + \omega_{pe}^2 \frac{\Omega}{k_y v_e}Z(\zeta_e^{\pm})
+ \omega_{pi}^2\frac{\Omega}{k v_{i,\parallel}^2}Z(\zeta_i^{\pm}) \nonumber \\ 
& & - \omega_{pi}^2\left(1 - \frac{T_{i\perp}}{T_{i,\parallel}}\right)
\left[1+\zeta_i^{\pm } Z(\zeta_i^{\pm})\right]\,,\label{eq:dispersion}
\end{eqnarray}
where  
$\Omega = \omega + i \gamma$ is the frequency of the instability, 
$k$ is the wavevector, 
$
\zeta_e^{\pm} = (\Omega \pm \Omega_{ce})/k_y v_e,
$
$
v_e = (2\,k_{\rm B}T_e/m_e)^{1/2},
$
$
\zeta_i^{\pm} = (\Omega \pm \Omega_{ci})/k_y v_{i,\parallel},
$
$
v_{i,\parallel} = (2\, k_{\rm B}T_{i,\parallel}/m_i)^{1/2}, 
$
and
$Z(\zeta)$ is the plasma dispersion function 
$$
Z(\zeta) = \frac{1}{\sqrt{\pi}}\int_{-\infty}^{\infty} dx \frac{\exp(-x^2)}{x-\zeta}~.
$$

To calculate the dispersion relation from Equation (\ref{eq:dispersion}), one possible choice for the set of parameters that we need to specify is
$
T_{i,\perp}/T_{i,\parallel}, T_{i, \parallel}, T_{e}, \beta_{e}, m_i/m_e.
$
As described in Section \ref{sec:setupmodel}, in the immediate post-shock downstream the values of $T_{i,\perp}/T_{i,\parallel}, T_{i, \parallel}, T_{e}, \beta_{e}$ can be derived from $M_s$ and $\beta_{p0}$. We adopt $m_i/m_e=49$ for the computations presented here, but we have checked that the dispersion relation for $m_i/m_e = 1836$ is almost identical.

\begin{figure*}[!b]
\begin{center}
\includegraphics[width=0.9\textwidth]{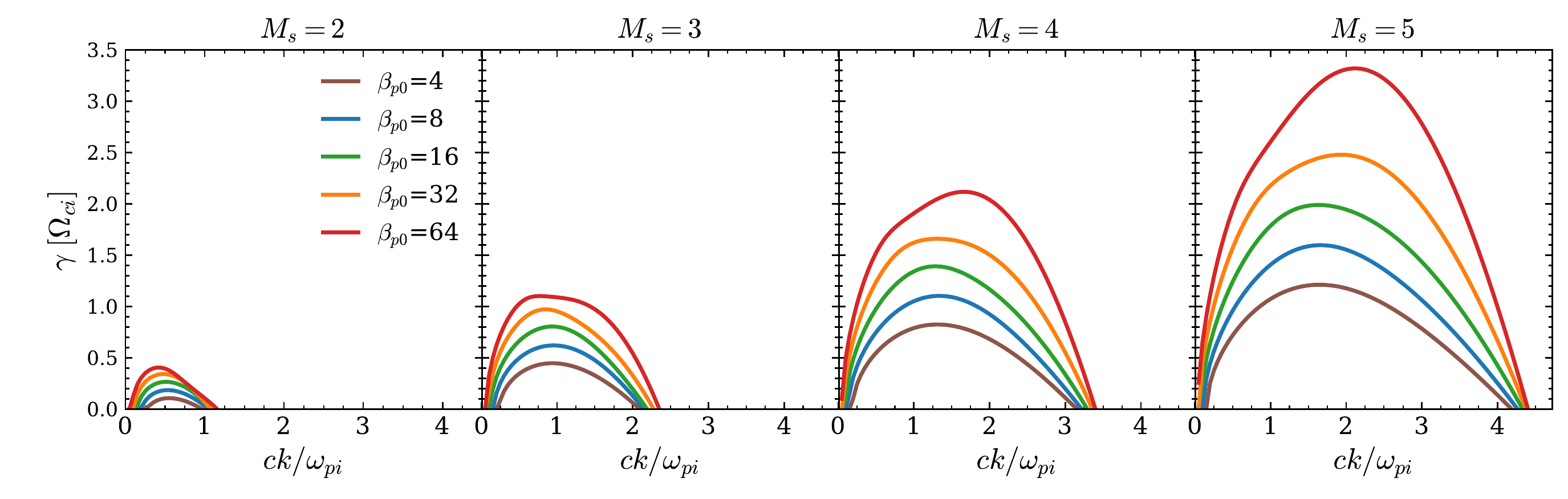}
\end{center}
\caption{Dependence of the dispersion relation (growth rate $\gamma$ as a function of wavevector $k$) of the proton cyclotron instability on the initial proton temperature anisotropy and plasma  beta, phrased in terms of the shock Mach number $M_s$ and the plasma beta $\beta_{p0}$. 
The relations of $M_s$ and $\beta_{p0}$ with the input parameters of the dispersion relation (i.e., $T_{i,\perp}/T_{i,\parallel}, T_{i, \parallel}, T_{e}, \beta_{e}$) can be derived as in Section \ref{sec:setupmodel}. 
}\label{fig:ICI}
\end{figure*}

Figure \ref{fig:ICI} shows the results. We see that with increasing $M_s$ (as indicated from the titles of panels, from left to right), which leads to an increasing $T_{i,\perp}/T_{i,\parallel}$, both the maximum 
 growth rate $\gamma$ and the wavevector $k$ of the fastest growing mode increase monotonically at fixed $\beta_{p0}$. 
At fixed  proton temperature anisotropy $T_{i,\perp}/T_{i,\parallel}$ (i.e., fixed $M_s$), the growth rate and the wavevector of the fastest growing mode increase moderately with $\beta_{p0}$ (as indicated by the different colors in each panel).

\bibliographystyle{apj}
\bibliography{Mendeley,heating}

\end{document}